\documentclass[11pt,letterpaper]{article}
\pdfoutput=1
\usepackage{jheppub}
\usepackage{color}
\usepackage{graphicx}
\usepackage{tabularx}
\usepackage{xspace}
\usepackage{tikz}

\usepackage{verbatim}
\usepackage{amsmath}
\usepackage{amssymb}
\usepackage[caption=false]{subfig}
\usepackage{url}
\usepackage{bbold}
\usepackage{slashed}
\usepackage{array}

\usepackage{multirow}
\usepackage{threeparttable}
\usepackage{paralist}
\usepackage{xspace}
\usepackage{upgreek}
\usepackage{lipsum}
\usepackage[utf8]{inputenc}

\newcommand\colvec[3][]{\begin{pmatrix}\ifx\relax#1\relax\else#1\\\fi#2\\#3\end{pmatrix}}

\newcommand{\beq}{\begin{equation}}
\newcommand{\beqn}{\begin{eqnarray}}
\newcommand{\eeq}{\end{equation}}
\newcommand{\eeqn}{\end{eqnarray}}
\newcommand\numberthis{\addtocounter{equation}{1}\tag{\theequation}}
\newcommand{\dbar}{\ensuremath{\mathchar'26\mkern-12mu d}}
\newcommand\order[1]{{\cal O}#1}

\DeclareRobustCommand{\Eq}[1]{Eq.~(\ref{#1})}

\DeclareRobustCommand{\Sec}[1]{Sec.~\ref{#1}}
\DeclareRobustCommand{\App}[1]{Appendix~\ref{#1}}
\DeclareRobustCommand{\Fig}[1]{Fig.~\ref{#1}}

\renewcommand{\vec}[1]{\mathbf{#1}}

\newcommand{\df}{\text{d}}

\begin{document}

\title{Principal Shapes and Squeezed Limits in the Effective Field Theory of Large Scale Structure}
\author{Daniele Bertolini \text{and}} 
\author{Mikhail P. Solon}

\affiliation{Berkeley Center for Theoretical Physics, University of California, Berkeley, CA 94720, USA }
\affiliation{Theoretical Physics Group, Lawrence Berkeley National Laboratory, Berkeley, CA 94720, USA}
\affiliation{Kavli Institute for the Physics and Mathematics of the Universe (WPI), University of Tokyo, Kashiwa 277-8583, Japan}
\emailAdd{dbertolini@lbl.gov}
\emailAdd{mpsolon@lbl.gov}

\date{\today}
\abstract{We apply an orthogonalization procedure on the effective field theory of large scale structure (EFT of LSS) shapes, relevant for the angle-averaged bispectrum and non-Gaussian covariance of the matter power spectrum at one loop. Assuming natural-sized EFT parameters, this identifies a linear combination of EFT shapes -- referred to as the principal shape -- that gives the dominant contribution for the whole kinematic plane, with subdominant combinations suppressed by a few orders of magnitude. For the covariance, our orthogonal transformation is in excellent agreement with a principal component analysis applied to available data. Additionally we find that, for both observables, the coefficients of the principal shapes are well approximated by the EFT coefficients appearing in the squeezed limit, and are thus measurable from power spectrum response functions. Employing data from N-body simulations for the growth-only response, 
we measure the single EFT coefficient describing the angle-averaged bispectrum with $\order(10\%)$ precision. These methods of shape orthogonalization and measurement of coefficients from response functions are valuable tools for developing the EFT of LSS framework, and can be applied to more general observables.}

\maketitle

\section{Introduction}
Current and upcoming surveys of large scale structure (LSS) will be instrumental in constraining the physics of the primordial universe and its expansion history (see, e.g., Refs.~\cite{2005astro.ph.10346T,2013Msngr.154...44J,2012SPIE.8446E..0ZM,2012arXiv1211.0310L,2013AJ....145...10D,2013arXiv1308.0847L,2015arXiv150804473D,
2011arXiv1106.1706S,Ellis:2012rn,2015arXiv150303757S,2013LRR....16....6A}).
In addition to the cosmic microwave background, late-time cosmological structures provide valuable information but require a precise and systematically improvable framework for incorporating the effects of gravitational nonlinear clustering. While N-body simulations have been a gold standard for describing nonlinear effects, an analytic approach to precision cosmology is imperative for a complementary understanding, especially of higher-point functions and primordial non-Gaussianities, baryonic and neutrino effects, and underlying uncertainties. 

Standard perturbation theory (SPT) constitutes the basic analytic framework (see, e.g., Ref.~\cite{2002PhR...367....1B} for a review), but lacks a proper description of the short scale dynamics and its feedback on the physics at large scales. The approach based on effective field theory (EFT) captures such effects through a general parametrization of the relevant physics at a given length scale, and thus allows for precise calculation of LSS observables, with controlled theoretical uncertainties~\cite{2012JCAP...07..051B,2014PhRvD..89d3521H,2012JHEP...09..082C,Foreman:2015lca,2016arXiv160200674B}. However, the parametrization introduces coefficients that need to be measured and that proliferate at higher orders in perturbation theory. While the appearance of these free coefficients is an unavoidable consequence of basic physical principles, it can undermine the utility of an analytic approach to LSS, given the already numerous nuisance parameters required to extract cosmological information. In this paper, we explore two simple strategies for addressing this challenge. 

The first strategy is to study correlations of EFT shapes and potentially identify a basis where a few shapes give the dominant contribution to observables. We refer to such shapes and their corresponding coefficients as principal EFT shapes and principal EFT coefficients. 
While EFT operators are linearly independent by construction, they may be highly correlated, especially for observables that depend only on a few kinematic variables. 
Hence, it is expedient to define a new basis where the shapes are maximally uncorrelated, and to then identify potential hierarchies among their contributions to the observable.
For example, the sum of two highly correlated shapes would dominate over their difference, granted the standard assumption of the EFT framework that coefficients are of the same order, 
regardless of the chosen basis. Depending on the uncertainties present in the theory and data, the subdominant shapes may or may not be included in the analysis.

The second strategy is to measure EFT coefficients appearing in higher $N$-point functions from responses of lower $N$-point functions to long-wavelength background fields. Beyond the power spectrum, the simulation of connected $N$-point functions becomes increasingly difficult and computationally expensive. They require a large number of modes to overcome cosmic variance, are more sensitive to systematic effects, and involve complicated estimators. On the other hand, we may 
access squeezed configurations of higher $N$-point functions and measure their EFT coefficients through the responses of lower $N$-point functions to background fields, which are more economical to compute.
Moreover, the nature of response functions as a derivative implies that a small number of realizations is sufficient for numerical simulation, and allows for a large fraction of the noise to cancel. 

While each of these strategies can be developed separately, in the present work, we focus on establishing an interesting connection between the two: principal EFT coefficients can be measured from response functions, i.e.,~EFT coefficients that appear in the squeezed limit may determine the dominant counterterm over the whole kinematic domain, even away from the squeezed region.
In particular, we consider squeezed limits of the angle-averaged bispectrum and non-Gaussian covariance of the matter power spectrum, and find that the principal EFT coefficient for these observables is well approximated by the EFT coefficient appearing when their squeezed limits are considered. For the case of the bispectrum, we measure the relevant EFT coefficient from N-body data for the first-order response corresponding to the squeezed configuration. For the case of the covariance, we compare our principal EFT shape to an actual principal component analysis (PCA) of the data and find excellent agreement, and we study the required anisotropic power spectrum response for measuring the principal coefficient from a squeezed configuration. 

The rest of the paper is organized as follows: in \Sec{sec:shapes}, we illustrate the rotation to a basis of uncorrelated EFT shapes, apply it to the angle-averaged bispectrum and non-Gaussian covariance of the power spectrum, and demonstrate how principal and squeezed coefficients align. In \Sec{sec:squeezed}, we provide the complete calculation of the one-loop angle-averaged squeezed bispectrum in SPT and EFT of LSS, and measure the squeezed bispectrum EFT coefficient from N-body data for the growth-only power spectrum response. We also derive the anisotropic power spectrum response corresponding to the squeezed covariance. Finally, we conclude in \Sec{sec:conclusion}. Throughout the paper we assume the reader is familiar with the basic SPT and EFT of LSS frameworks; a brief review can be found in \App{app:EFT}. In Appendices \ref{app:shapes} and \ref{app:OneLoopBs}, we collect calculational details of the EFT shapes and of the one-loop squeezed bispectrum, respectively.

\section{Orthogonal EFT Shapes}
\label{sec:shapes}
We begin by introducing the notion of orthogonality among EFT contributions to LSS observables. 
This allows to define a basis where the shapes are maximally uncorrelated and identify principal EFT shapes. 
A brief review of the EFT framework and further details of the calculations in this section can be found in Appendices \ref{app:EFT} and \ref{app:shapes}.	

In the EFT of LSS, the feedback of short-scale modes on the dynamics of large-scale modes is parametrized by a stress tensor $\uptau^{ij}$. The stress tensor is constructed by including all mode-coupling functions compatible with the symmetries of the system, and associating with each function an arbitrary coefficient $\bar{c}$. Following Ref.~\cite{Bertolini:2016bmt}, the stress tensor through next-to-next-to-leading order (NNLO) in fields and at leading order in the derivative expansion is given by
\begin{align*}\label{eq:stresstensor1}
&k_i { \uptau}^{ij}  = \, \bar{c}_s^\delta k^j  \delta({\vec k})
+ \int \df\vec{q} \sum_{n=1}^3 \bar{c}_n^{\delta\delta}  \delta({{\vec q}})\delta(\vec{k} - {{\vec q}}) k_i e^{ij}_n (\vec{q}, \vec{k} - \vec{q})\\
&\quad\quad\quad + \int \df\vec{q} \sum_{n=2}^3 {\bar{c}_n^{\theta\theta}\over \mathcal{H}^2f^2}  \theta({{\vec q}})\theta(\vec{k} - {{\vec q}})) k_i e^{ij}_n (\vec{q}, \vec{k} - \vec{q})\\
&\quad\quad\quad  +  \int \df\vec{q}_1 \df\vec{q}_2  \sum_{n=1}^{6} \bar{c}_n^{\delta\delta\delta}   \delta({{\vec q}_1}) \delta({{\vec q}_2}) \delta(\vec{k} - {{\vec q}_1} - \vec{q}_2)  k_i E^{ij}_n (\vec{q}_1, \vec{q}_2, \vec{k} - {{\vec q}_1} - \vec{q}_2) \,, \numberthis 
\end{align*} 
where the coefficients have dimension $[k]^{-2}$, and their time dependence matches the SPT loop's. That is, each coefficient takes the form $\bar{c}=[\mathcal{H}f(\tau)D(\tau)]^2c$, where $c$
is time independent, $D(\tau)$ is the linear growth function and $f(\tau)=1/\mathcal{H}\,\df\log D(\tau)/\df\tau$.
The functions $e_n^{ij}$ and $E_n^{ij}$ are specified in Appendix~\ref{app:EFT}.

For an LSS observable at one loop, the contributions from these mode-couplings are linear in the coefficients and may be written as
\beq\label{eq:observable}
O(\kappa) = O_{\rm SPT}(\kappa) + \sum_i \bar{c}_i S_i(\kappa) \,,
\eeq
where $\kappa$ collectively denotes kinematic variables.
The first term includes the tree-level and one-loop contributions from SPT, while the second term includes the EFT corrections, and is written as a sum over EFT coefficients $\bar{c}_i$, which are linear combinations of the coefficients appearing in \Eq{eq:stresstensor1}, with associated shapes $S_i$. 
Note that we have defined these shapes in terms of the observable, as opposed to mode-coupling functions appearing in the stress tensor or shapes in the EFT kernels. In particular, the shapes appearing in the stress tensor are linearly independent by construction, but the shapes $S_i$ could become highly correlated (even redundant) depending on the observable. While it is straightforward to identify and remove linear dependence, the notion of correlation or similarity among shapes is more subtle and can be quantified by defining a dot product.
 
Let us write the dot product between two shapes $S_i$ and $S_j$ as
\beq\label{eq:dot}
S_i \cdot S_j  = \int {\rm d} \kappa \, S_i (\kappa) \, S_j (\kappa) W (\kappa) \,,	
\eeq
where the integration is over the kinematic variables, and $W$ is a weighting function. Note that $0 \leq | S_i \cdot S_j | < \infty$ since we assume the integration is over a bounded region, within which the shapes and weighting function are well-defined. In an analysis with data, the weighting function may be chosen to incorporate uncertainties to better inform whether the difference between two shapes is detectable. For now, we simply aim to explore the correlations among EFT shapes, and take $W=1$.

The dot product in \Eq{eq:dot} defines a measure of the correlation between two shapes, so that orthogonal shapes are uncorrelated according to this measure.
For obtaining a basis of uncorrelated shapes, we consider the matrix of dot products, $M_{ij} =  {S}_i \cdot {S}_j$, and collect its eigenvectors in an orthogonal matrix $\boldsymbol{U}$ such that $\boldsymbol{U} \boldsymbol{M} \boldsymbol{U}^T = \boldsymbol{M}^\prime$ is diagonal. 
The basis of orthogonal shapes and their respective coefficients is then given by ${S}^\prime_i = U_{ij} {S}_j$
and $\bar{{c}}^\prime_i = U_{ij} \bar{c}_j$.
Note that for studying the correlations among shapes independent of their overall scalings, it is convenient to define normalized shapes ${\hat S}_i = S_i / \sqrt{S_i \cdot S_i}$ and compute the correlation matrix ${\hat M}_{ij} =  {\hat S}_i \cdot {\hat S}_j$. 

Let us look at the implications of this 
rotation for two examples: the angle-averaged bispectrum and the non-Gaussian covariance of the matter power spectrum.   
For the numerical results of the next two sections, we assume the following cosmology: $\Omega_m = 0.286$, $\Omega_b = 0.047$, $\text{h}=0.7$, $n_s = 0.96$, $\sigma_8 = 0.82$.

\subsection{Angle-Averaged Bispectrum}
\label{sec:bispectrum}
\begin{figure*}[t]
\begin{center}
\includegraphics[width=11cm]{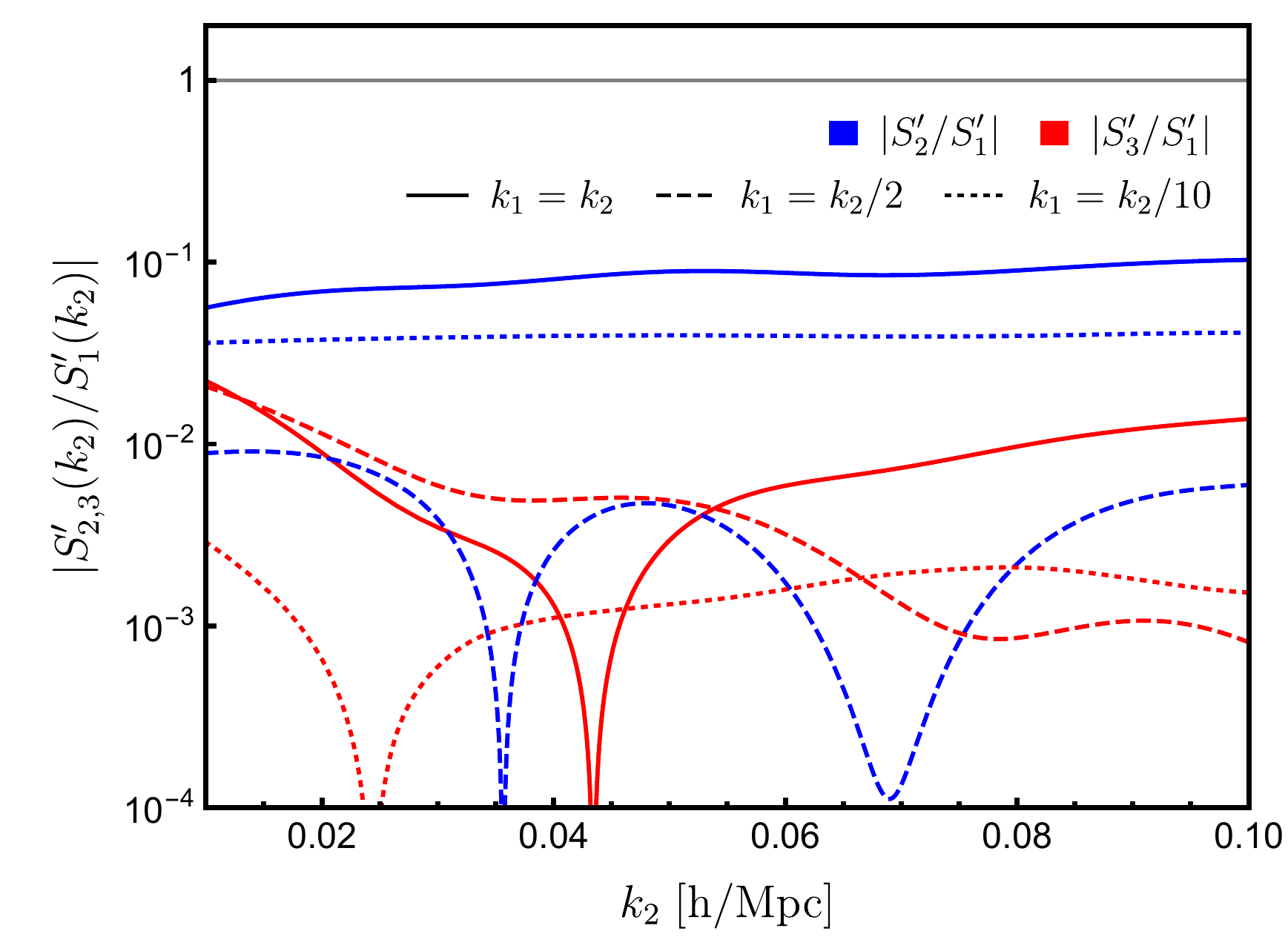}
\includegraphics[width=11cm]{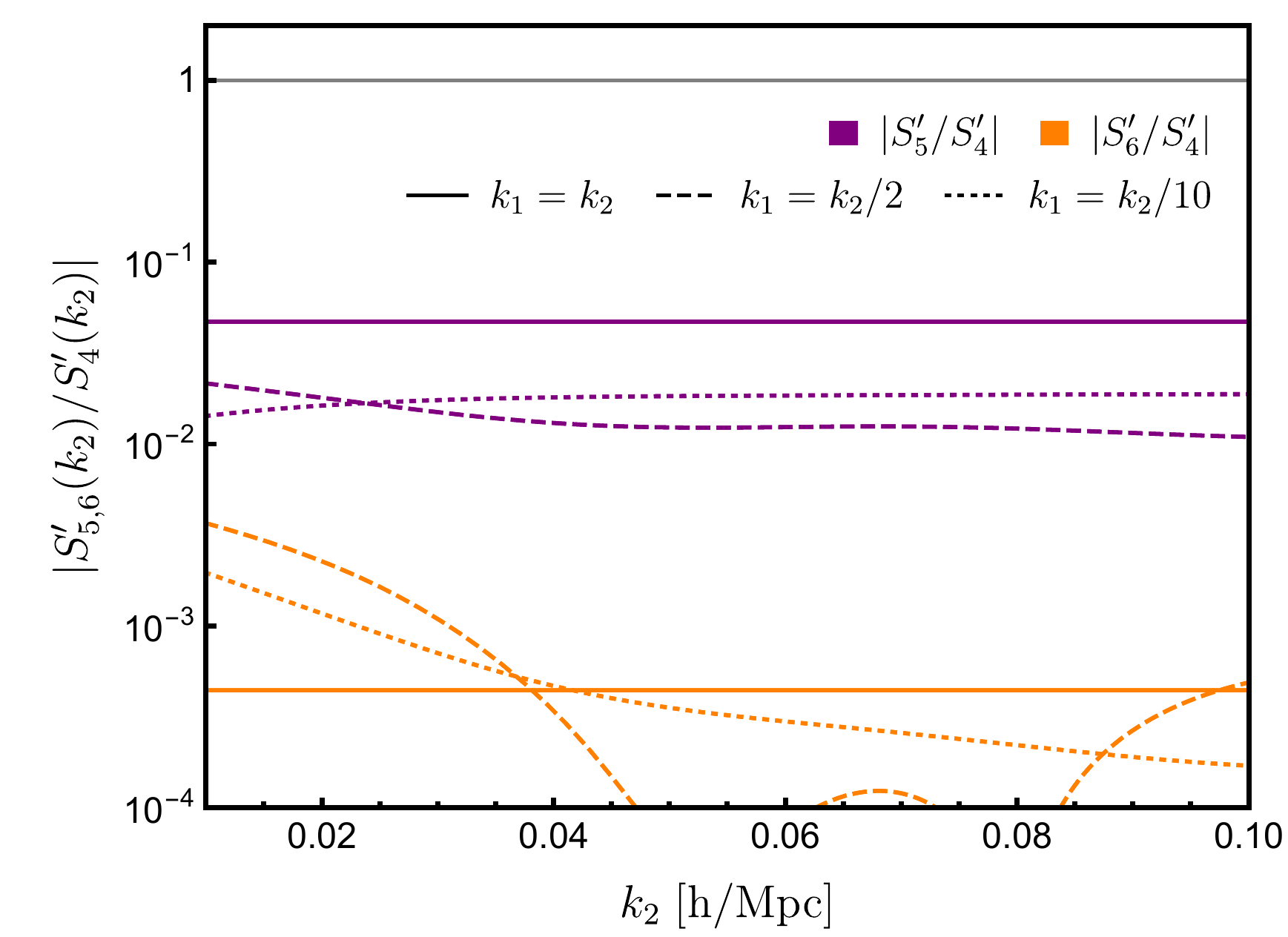}
\caption{Top: the shapes in the uncorrelated basis for the angle-averaged bispectrum, expressed as $|S_{i}^\prime/S_{1}^\prime|$ for $i=2,3$ (blue, red), where $S_{1}^\prime$ is the principal shape. Bottom: the shapes in the uncorrelated basis for the covariance, expressed as $|S_{i}^\prime/S_{4}^\prime|$ for $i=5,6$ (purple, orange), where $S_{4}^\prime$ is the principal shape. The solid, dashed and dotted lines have $k_1=k_2$, $k_1 = k_2/2$ and $k_1=k_2/10$, respectively. The latter is representative of the squeezed limit, $k_1\ll k_2$.}
\label{fig:bishapes}
\end{center}
\end{figure*}

The EFT contributions to the angle-averaged bispectrum are counterterm diagrams involving the kernels ${\widetilde F}_1$ and ${\widetilde F}_2$, and depend on the coefficients $\bar{c}_{s,1,2,3}$. Note that these coefficients are related to those appearing in \Eq{eq:stresstensor1} through the redefinitions specified in \App{app:EFT}.
To simplify the analysis in this example, let us consider $\bar{c}_s$ fixed (e.g., known from the power spectrum), and apply the orthogonalization on the subspace of shapes corresponding to coefficients $\bar{c}_{1,2,3}$.
The relevant EFT contributions are then contained in the diagram involving the EFT kernel ${\widetilde F}_2$:
\beq\label{eq:shapesBi}
\begin{tikzpicture}[baseline={([yshift=-.5ex]current bounding box.center)}]
\draw[thick,dashed] (-0.5,0) -- (0,1);
\draw[ thick,dashed] (0.5,0) -- (0,1);
\draw [fill=lightgray] (-0.09,1.09) rectangle (0.09,0.91);
\draw [fill=black] (-0.5,0) circle (2pt);
\draw [fill=black] (0.5,0) circle (2pt);
 \end{tikzpicture} \, \supset \, \sum_{i=1}^{3} \bar{c}_i \, S_i (k_1,k_2) \, ,
\eeq
where the explicit expressions for $S_{1,2,3}$ can be found in \App{app:shapes}.
Considering kinematics in the linear to mildly nonlinear regime, let us, for example, take $\int \df \kappa = \int_{0.01}^{0.1} \df k_1 \int_{0.01}^{0.1}  \df k_2$.  In the basis $\{ {\hat S}_1, {\hat S}_2, {\hat S}_3 \}$, the correlation matrix and its eigenvalues are
\begin{align}\label{eq:bicor}
\boldsymbol{\hat M} = \left(
\begin{array}{ccc}
 1 & 0.998 & 0.992 \\
 0.998 & 1 & 0.998 \\
 0.992 & 0.998 & 1 \\
\end{array}
\right) \,, \quad \{3\,, 8 \times 10^{-3} \,,  5 \times 10^{-5} \} \,.
\end{align}
The form of the above matrix and its hierarchical eigenvalues indicate that the shapes are highly correlated and that the shapes in the orthogonal basis have large hierarchies. In the basis $\{ {S}_1, {S}_2, {S}_3 \}$, the orthogonal transformation is given by
\begin{align}\label{eq:birot}
\boldsymbol{U} =  
\left(
\begin{array}{ccc}
 -0.825 & -0.511 & -0.242 \\
 0.530 & -0.547 & -0.648 \\
 0.199 & -0.663 & 0.722 \\
\end{array}
\right) \,,
\end{align}
and the orthogonal shapes $S_i^\prime(k_1,k_2)$, normalized to the dominant shape $S_1^\prime(k_1,k_2)$, are shown in the top panel of Fig~\ref{fig:bishapes}. Assuming natural sizes for the coefficients $\bar{c}^\prime_i$,\footnote{\label{foot:num}The EFT coefficients have natural size $\order(k_{\rm NL}^{-2}) = \order(10 \, \rm{Mpc}^2/\rm{h}^2)$. For example, taking $\{ \bar{c}_1, \bar{c}_2, \bar{c}_3\} = \{18.5 \,, -41.1 \,, 62.4  \}$ Mpc$^2$/h$^2$ from Ref.~\cite{2015JCAP...05..007B},  we find $\{ \bar{c}_1^\prime, \bar{c}_2^\prime, \bar{c}_3^\prime\} = \{ 9.4\,,-8.1\,,76.0 \}$ Mpc$^2$/h$^2$.
} the hierarchy among shapes implies that the contributions to the angle-averaged bispectrum from $S^\prime_{2,3}$, relative to $S^\prime_1$, are at most $\order(10\%)$, depending on the ratio $k_2/k_1$.

\subsection{Covariance}
\label{sec:shapesCov}
The analysis for the non-Gaussian covariance of the matter power spectrum follows in a similar manner. As derived in Ref.~\cite{2015arXiv151207630B}, there are three new EFT operators in addition to the power spectrum and bispectrum counterterms, whose coefficients are denoted as  $\bar{c}_{4,5,6}$. Again, these are related to the coefficients appearing in \Eq{eq:stresstensor1} by the redefinitions specified in \App{app:EFT}. The contribution of these three new operators is contained in the  diagram involving the EFT kernel ${\widetilde F}_3$:
\beq\label{eq:shapesCov}
\begin{tikzpicture}[baseline={([yshift=-.5ex]current bounding box.center)}]
\draw[thick,dashed] (-0.6,0) -- (0,1);
\draw[ thick,dashed] (0.6,0) -- (0,1);
\draw[ thick,dashed] (0,-0.15) -- (0,1);
\draw [fill=lightgray] (-0.09,1.09) rectangle (0.09,0.91);
\draw [fill=black] (-0.6,0) circle (2pt);
\draw [fill=black] (0.6,0) circle (2pt);
\draw [fill=black] (0,-0.15) circle (2pt);
 \end{tikzpicture} \, \supset \, \sum_{i=4}^{6} \bar{c}_i \, S_i (k_1,k_2) \, ,
\eeq
where the explicit expressions for $S_{4,5,6}$ can be found in \App{app:shapes}.
Using the same kinematic region as in \Sec{sec:bispectrum}, and working in the basis $\{ {\hat S}_4, {\hat S}_5, {\hat S}_6 \}$, the correlation matrix and its eigenvalues are given by
\begin{align}\label{eq:covM}
\boldsymbol{\hat M} = \left(
\begin{array}{ccc}
 1 & 1 & -0.991 \\
1 & 1 & -0.987 \\
 -0.991 & -0.987 & 1 \\
\end{array}
\right)\,, \quad \{3\,, 10^{-2} \,, 2 \times 10^{-5} \} \, .
\end{align}
These again indicate that the shapes are highly correlated, and that the shapes in the orthogonal basis have large hierarchies. In the basis $\{ {S}_4, {S}_5, {S}_6 \}$, the orthogonal transformation is given by
\begin{align}\label{eq:covrot}
\boldsymbol{U} = \left(
\begin{array}{ccc}
 0.132 & 0.043 & -0.99 \\
 0.923 & 0.359 & 0.138 \\
 0.362 & -0.932 & 0.008 \\
\end{array}
\right) \,,
\end{align}
and the corresponding shapes $S_i^\prime(k_1,k_2)$, normalized to the dominant shape $S_4^\prime(k_1,k_2)$, are shown in the bottom panel of Fig~\ref{fig:bishapes}. 

In Ref.~\cite{2015arXiv151207630B}, a PCA was applied to covariance data from N-body simulations, yielding the transformation
\begin{align}\label{eq:covPCA}
\boldsymbol{U}_{\rm PCA} =  \left(
\begin{array}{ccc}
 0.121 & 0.038 & -0.992 \\
 0.926 & 0.355 & 0.126 \\
  0.357 & -0.934 & 0.008 \\
\end{array}
\right)\,.
\end{align}
The agreement between Eqs.~(\ref{eq:covrot}) and~(\ref{eq:covPCA}) is remarkable, and can be improved, e.g., by using a dot product that incorporates the same kinematic domain and wavenumber binning as the data analysis. 

We find that our results in Eqs.~(\ref{eq:bicor}) and~(\ref{eq:birot}) for the bispectrum and Eqs.~(\ref{eq:covM}) and~(\ref{eq:covrot}) for the covariance are not sensitive to the choice of the integration range, since the shapes $S_{i}(k_1,k_2)$ are highly correlated in the whole kinematic plane. For future work, it would be interesting to study less degenerate observables and include realistic constraints on the experimentally relevant kinematic domain, as these might break some degeneracies and lead to multiple principal shapes (see, e.g., Ref.~\cite{Welling:2016dng} for a study of correlations in measurements of primordial non-Gaussianities).

\subsection{Alignment of Principal and Squeezed EFT Coefficients}
\label{subsec:alignment}
The previous examples clearly illustrate the utility of employing an uncorrelated basis
for analyzing the EFT contributions to LSS observables. There may be large hierarchies among the EFT contributions, and thus certain shapes may be neglected depending on the theoretical and data uncertainties present in an analysis. Conversely, the same method informs the level of backgrounds required for detecting the subdominant shapes, and for which wavenumbers their signals peak.

Let us now turn to an interesting connection between the principal EFT coefficients and the squeezed limits of the observables. Note from Fig.~\ref{fig:bishapes} that the hierarchy among shapes is consistent in the whole kinematic plane, including the squeezed limit, $k_1 \ll k_2$ (dotted curves). This implies that the linear combination of coefficients in the squeezed limit should align well with the principal EFT coefficient.
Indeed, the normalized linear combinations of coefficients in the squeezed limit may be written as
\begin{align}\label{eq:align}
\bar{c}_b &= {-3\bar{c}_1 -2 \bar{c}_2 -\bar{c}_3 \over \sqrt{14}} = 0.999 \bar{c}_1^\prime + 0.041\bar{c}_2^\prime + 0.002 \bar{c}_3^\prime \,, \nonumber \\
\bar{c}_t &= {48 \bar{c}_4  + 16  \bar{c}_5 - 315 \bar{c}_6 \over \sqrt{101785}} = 0.9998 \bar{c}_4^\prime + 0.0201 \bar{c}_5^\prime -  0.0003 \bar{c}_6^\prime \,,
\end{align}
where $\bar{c}_b$ and $\bar{c}_t$ are for the angle-averaged bispectrum and covariance, respectively. The results in terms of the coefficients $\bar{c}^\prime_i$ follow from Eqs.~(\ref{eq:birot}) and~(\ref{eq:covrot}).

Assuming natural sizes for the coefficients $\bar{c}^\prime_i$ (see, e.g., footnote \ref{foot:num}), the principal EFT coefficients $\bar{c}_1^\prime$ and $\bar{c}_4^\prime$ are thus well approximated by $\bar{c}_b$ and $\bar{c}_t$, respectively.
As discussed in the previous section, the dominance of a single EFT operator for the covariance was also found in Ref.~\cite{2015arXiv151207630B}, by applying a PCA to N-body simulation data. The PCA combination, whose coefficient was labeled $c^*$ in Ref.~\cite{2015arXiv151207630B}, essentially coincides with the dominant shape, i.e.,~$\bar{c}'_4\approx \bar{c}_t\approx -c^*$.

Equation~(\ref{eq:align}) implies that the principal EFT coefficients may be measured from the response functions corresponding to the squeezed observables. 
In the next section, we consider these squeezed observables and their relation to response functions in detail. In particular, $\bar{c}_b$ can be measured from the growth-only response $G_1(k)$, while we find that $\bar{c}_t$ requires responses to anisotropic backgrounds.
The measurement of $\bar{c}_b$ from N-body data is presented in Sec~\ref{subsec:measure}.

\section{Squeezed Angle-Averaged Bispectrum and Covariance}
\label{sec:squeezed}
Squeezed limits of matter density correlators, where one or more external wavenumbers are taken to be soft, have been studied extensively in the literature (see, e.g., Refs.~\cite{Kehagias:2013yd,Peloso:2013zw,Creminelli:2013mca,Peloso:2013spa,Creminelli:2013poa,Kehagias:2013rpa,Valageas:2013cma,
Valageas:2013zda,Kehagias:2013paa,Creminelli:2013nua,Nishimichi:2014jna,Ben-Dayan:2014hsa,Horn:2015dra,Dai:2015jaa,Sherwin:2012nh,Takada:2013bfn,Li:2014sga,Wagner:2015gva}). In particular, 
squeezed limits of higher-order correlators
can be related to responses of lower order correlators to long-wavelength background fields. For example, as shown in Ref.~\cite{Wagner:2015gva}, the angle-averaged squeezed limit of 
 the connected matter $N$-point function
is related to the response of the matter power spectrum to long-wavelength isotropic density perturbations as
\begin{align}\label{eq:Rn}
 R_{N-2}(k) \equiv \lim_{q_i\to 0}\frac{\big\langle \Gamma(\vec{k},\vec{k}',\vec{q}_1,...,\vec{q}_{N-2})\big\rangle}{P(k)P_L(q_1)...P_L(q_{N-2})}=\frac{1}{P(k)}\frac{\df^{N-2}P(k|\delta_L)}{\df\delta_L^{N-2}}\Bigg|_{\delta_L=0} \, .
\end{align}
Here, $\Gamma(\vec{k},\vec{k}',\vec{q}_1,...,\vec{q}_{N-2}) \equiv \langle\delta(\vec{k})\delta(\vec{k}')\delta(\vec{q}_1)...\delta(\vec{q}_{N-2})\rangle_c$ is the connected matter $N$-point function, and the angle brackets $\langle  \ \rangle$ in \Eq{eq:Rn} denote an average over the directions of the squeezed wavevectors $\vec{q}_i$. We have denoted the power spectrum in the presence of the long-wavelength background fields $\delta_L(\vec{q}_i)$ by $P(k|\delta_L)$, and kept the time dependence implicit.

The responses $R_{N-2}(k)$ are properties of the power spectrum that contain information about higher-order $N$-point functions, which are otherwise challenging to measure directly.
At the same time, these responses are relatively clean and economical to compute using, e.g., finite difference methods where a large fraction of the sample variance cancels. Beyond Eq.~(\ref{eq:Rn}), more general relations may also be constructed, such as responses to anisotropic backgrounds and responses of the bispectrum, and would provide even more information on higher-order correlators.
For these reasons, response functions are valuable for precise determination of EFT coefficients.

Aside from these advantages, we are interested in measuring the EFT coefficients in the squeezed limit since, as discussed in Sec.~\ref{subsec:alignment}, for the cases of the angle-averaged bispectrum and covariance they align remarkably well with the principal EFT coefficients. In this section, we present results for the one-loop angle-averaged squeezed bispectrum in SPT and EFT of LSS, and measure the corresponding EFT coefficient from N-body simulation data. For the covariance, the required response involves anisotropic backgrounds, and we generalize Eq.~(\ref{eq:Rn}) accordingly. We present tree-level results for the set of responses, and provide the EFT counterterms that would be measured if data becomes available.

\subsection{Squeezed Angle-Averaged Bispectrum}
\begin{figure}
\begin{center}
\includegraphics[width=11cm]{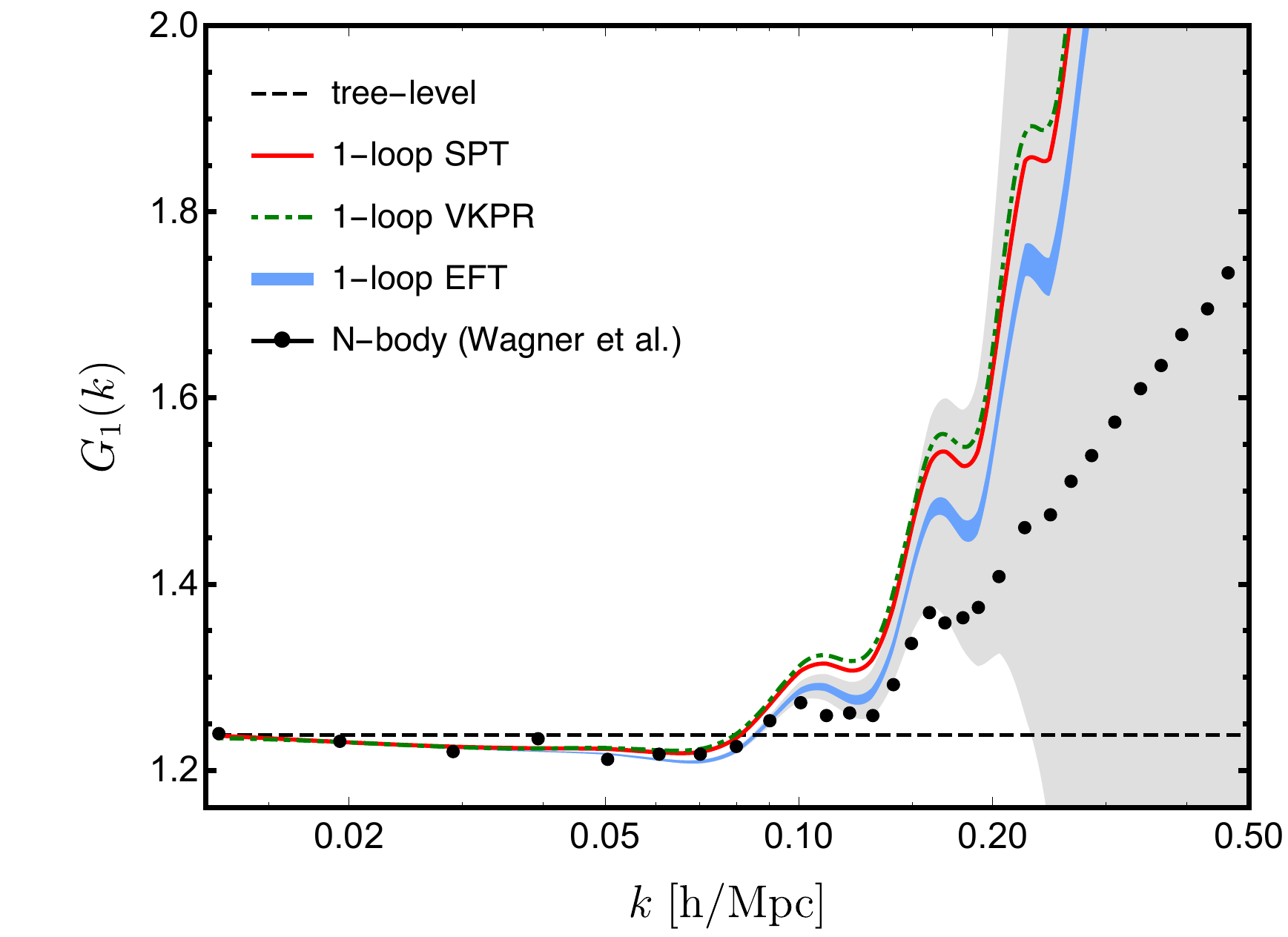}
\caption{The growth-only response $G_1(k)$ for tree-level (black dashed), one loop SPT (red solid), the VKPR ansatz (green dot-dashed), one loop EFT (blue band), and N-body simulation data from Ref.~\cite{Wagner:2015gva} (black points). The thickness of the blue and gray bands represents the uncertainty in the measured EFT coefficient $c_E$ and the higher-order perturbative corrections, respectively. The data points include error bars that are smaller than the plot markers.}\label{fig:G1}
\end{center}
\end{figure}

For the case of the angle-averaged squeezed bispectrum, \Eq{eq:Rn} reads
\begin{align}\label{eq:R1}
R_1(k)\equiv\lim_{q\to 0}\,\frac{\langle B(\vec{k},\vec{k}',\vec{q})\rangle}{P(k)P_L(q)}=\frac{1}{P(k)}\frac{\df P(k|\delta_L)}{\df\delta_L}\Bigg|_{\delta_L=0}  \, .
\end{align}
The function $R_1(k)$ can be measured as the (linear) response of the matter power spectrum to an isotropic long-wavelength density perturbation, employing, e.g., the separate universe picture~\cite{Wagner:2014aka}. In this approach, the isotropic long-wavelength background is reabsorbed into a modified cosmology with a non-zero curvature, and we may thus factor out the contributions to $R_1(k)$ due to the remapping of the background density and comoving coordinates from the modified to the fiducial cosmology. This isolates the growth-only part of the response,
\begin{align}\label{eq:G1}
G_1(k)= R_1(k)-1+\frac{1}{3}\frac{\df\log P(k)}{\df\log k} \, ,
\end{align}
which describes the change in the growth of the short modes due to the presence of the long-wavelength background.
The function $G_1(k)$ has been measured from N-body simulations in Ref.~\cite{Wagner:2015gva}, and we provide here the one-loop SPT prediction, including the EFT counterterms. At tree level, one simply finds
\begin{align}\label{eq:G1tree}
G_1^\text{tree}(k)=\lim_{q\to 0}\,\frac{\langle B_\text{tree}(\vec{k},\vec{k}',\vec{q})\rangle}{P_L(k)P_L(q)}-1+\frac{1}{3}\frac{\df\log P_L(k)}{\df\log k}=\frac{26}{21}.
\end{align}
The SPT one-loop correction is given by
\begin{align}\label{eq:G1loop}
G_1^\text{1-loop}(k)=\lim_{q\to 0}\,\frac{\langle B_\text{1-loop}(\vec{k},\vec{k}',\vec{q})\rangle}{P_L(k)P_L(q)}-\frac{P_\text{1-loop}(k)}{P_L(k)}\left[\frac{47}{21}-\frac{1}{3}\frac{\df\log P_\text{1-loop}(k)}{\df\log k}\right],
\end{align}
where $B_\text{1-loop}(\vec{k},\vec{k}',\vec{q})$ and $P_\text{1-loop}(k)$ are the one-loop contributions to the bispectrum and power spectrum, respectively. Of the four one-loop bispectrum diagrams (shown in Appendix~\ref{app:BsSPT}), only $B_{411}, B_{321}^{a}$ and $B_{321}^{b}$ contribute to the squeezed limit,
 with diagram $B_{222}$ suppressed by additional powers of $q/k$. We collect the 
expressions for $B_{411}, B_{321}^{a}$ and $B_{321}^{b}$ in \App{app:BsSPT}, and show the full one-loop SPT result for $G_1(k)$ in \Fig{fig:G1}.

The EFT diagrams 
and the corresponding amplitudes are collected in \App{app:BsEFT}.
In the squeezed limit, we are left with one counterterm for the response,
\begin{align}\label{eq:G1eft}
G_1^\text{EFT}(k)=-\frac{136}{2079}\bar{c}_sk^2+\frac{8\sqrt{14}}{99}\underbrace{\frac{-3\bar{c}_1-2\bar{c}_2-\bar{c}_3}{\sqrt{14}}}_{\bar{c}_b} \, k^2,
\end{align}
where $\bar{c}_b$ is the normalized combination of $\bar{c}_{1,2,3}$ introduced in Eq.~(\ref{eq:align}).
Assuming that $\bar{c}_s$ is known, we can measure $\bar{c}_b$ from N-body data, and thus determine the principal EFT shape that renormalizes the angle-averaged bispectrum in the whole kinematic plane, up to corrections of $\mathcal{O}(10\%)$. 

Before moving on to the measurement of $\bar{c}_b$ in the next section, let us note that  
Valageas~\cite{Valageas:2013zda} as well as Kehagias, Perrier and Riotto~\cite{Kehagias:2013paa} have proposed an ansatz for $G_1(k)$ based on the form of the linear growth function in the curved background.
 In the VKPR ansatz, $G_1(k)$ can be written in terms of the power spectrum of the fiducial cosmology at all orders in perturbation theory (see Appendix~\ref{app:VKPR} for a brief review). On the other hand, our explicit calculation of the one-loop correction in Eq.~(\ref{eq:G1loop}) does not factorize in this manner, and accordingly, the EFT contribution in Eq.~(\ref{eq:G1eft}) does not depend on only $\bar{c}_s$, but also on $\bar{c}_{1,2,3}$. Nonetheless, assuming the ansatz, we may also measure the EFT coefficient $\bar{c}_s$ and compare to previous measurements as a consistency check. 

\subsection{Measurement of $\bar{c}_b$}
\label{subsec:measure}
Having specified the one-loop SPT prediction in Eqs.~(\ref{eq:G1tree}) and~(\ref{eq:G1loop}), as well as the EFT contribution in Eq.~(\ref{eq:G1eft}), we may now measure the EFT coefficient $\bar{c}_b$ from N-body simulations. We employ data for the growth-only response from Wagner et al.~\cite{Wagner:2015gva}, whose cosmological parameters and simulation volume are $\Omega_m = 0.27$, $\Omega_b \text{h}^2 = 0.023$, $\text{h}=0.7$, $n_s = 0.95$, $\mathcal{A}_s = 2.2\times 10^{-9}$, and $V=(500 \, {\rm Mpc/h})^3$.

To detect the EFT contribution of the form $\sim k^2$, let us define the coefficient estimator $c_E$ and its uncertainties $\Delta c_E$ as 
\begin{align}
\label{eq:E}
c_E = { G_1^{\rm data} - G_1^{\rm tree} - G_1^\text{1-loop}  \over k^2} \,, \quad \Delta c_E = {\sqrt{ \left( \Delta G_1^{\rm data} \right)^2
+ \left( \Delta G_1^{\rm theory} \right)^2 } \over k^2} \,,
\end{align}
and perform a fit using the standard least squares method. Here, $\Delta G_1^{\rm data}$ is the uncertainty on the data, and $\Delta G_1^{\rm theory}$ is the theoretical uncertainty from missing higher-order SPT and EFT contributions. In our analysis, we neglect the correlations between different $k$ values. Ideally, we would measure the EFT parameters from data at small wavenumbers ($k \ll k_{\rm NL}$), where higher-order corrections are suppressed ($\Delta G_1^{\rm theory} \approx 0$). However, while $c_E$ is a constant as $k \to 0$, the uncertainty $\Delta c_E$ increases as $1/k^2$ and moreover $\Delta G_1^{\rm data}$ increases due to residual sample variance effects in the simulation. On the other hand, data at large wavenumbers should be down-weighted to avoid fitting to signal from higher-order effects. This can be done in a systematic way that maximizes the available information by specifying $\Delta G_1^{\rm theory}$, as opposed to, e.g., the common practice of imposing a blunt fitting window. 

From the expression for $G_1(k)$ in Eq.~(\ref{eq:G1}), expanded order by order in both SPT and EFT perturbations, we may write
\begin{align}\label{eq:dGtheory}
\left( \Delta G_1^{\rm theory} \right)^2 &= \sum_{n=1}^\infty \left[ \alpha_{\rm SPT}^{2(n+1)}  + \alpha_{\rm EFT}^{2(n+1)} +   \alpha_{\rm SPT}^{2n}  \sum_{m=1}^\infty \alpha_{\rm EFT}^{2m} \right] \, ,
\end{align}
where we have assumed that the $n$-loop SPT and EFT corrections respectively scale as $\alpha^{n}_{\rm SPT}$ and $\alpha^{n}_{\rm EFT}$ relative to the tree-level result.
In this estimate for the uncertainties, we neglect $\order(1)$ factors and add terms in quadrature. The size of EFT corrections follows from the power counting used in constructing the stress tensor. For SPT, we may estimate it from representative higher-order terms. The one-loop SPT corrections to $G_1(k)$ involve $B_\text{1-loop} / P_{L}$ and $P_\text{1-loop} / P_{L}$, while the two-loop corrections involve $B_\text{2-loop} / P_{ L}$, $P_\text{2-loop} / P_{L}$, $B_\text{1-loop} P_\text{1-loop} / P_{L}^2$ and $P_\text{1-loop} ^2 / P_{L}^2$. 
We thus take as default
\begin{align}
\alpha_{\rm EFT}^{} = {k^2 \over k_{\rm NL}^2 }  \,, \quad \alpha_{\rm SPT}^{} =   {P_\text{1-loop} \over P_{ L} }   \,.
\end{align}
We have checked that these are consistent with estimates assuming a scaling universe (see, e.g., Ref.~\cite{Pajer:2013jj}); further checks may be performed using explicit results for $B_\text{2-loop}$ and $P_\text{2-loop}$~\cite{2014arXiv1406.4143A,2015JCAP...05..007B,2014JCAP...07..057C,Baldauf:2015aha}. 

\begin{figure*}[t]
\begin{center}
\includegraphics[width=11cm]{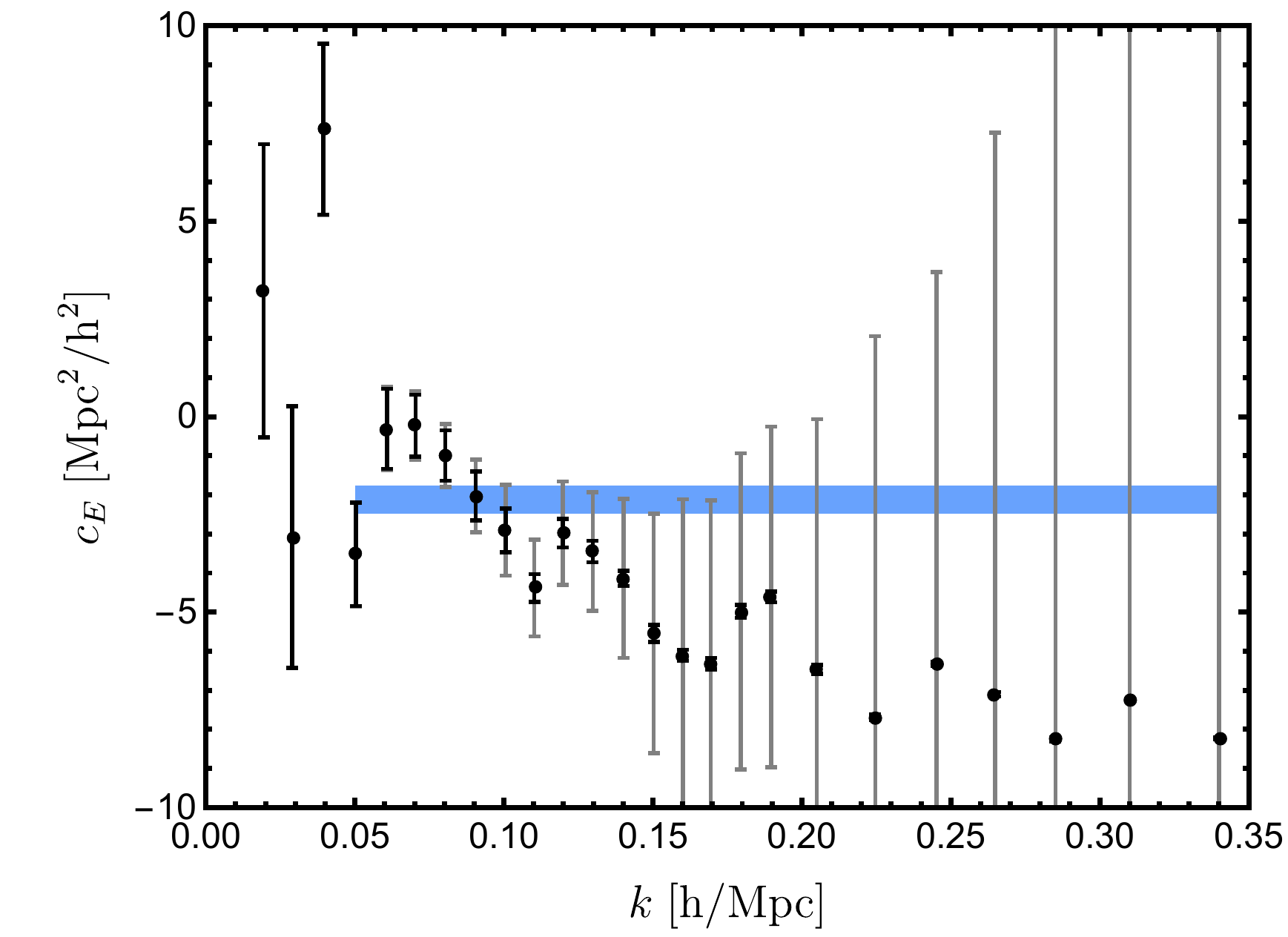}
\includegraphics[width=11cm]{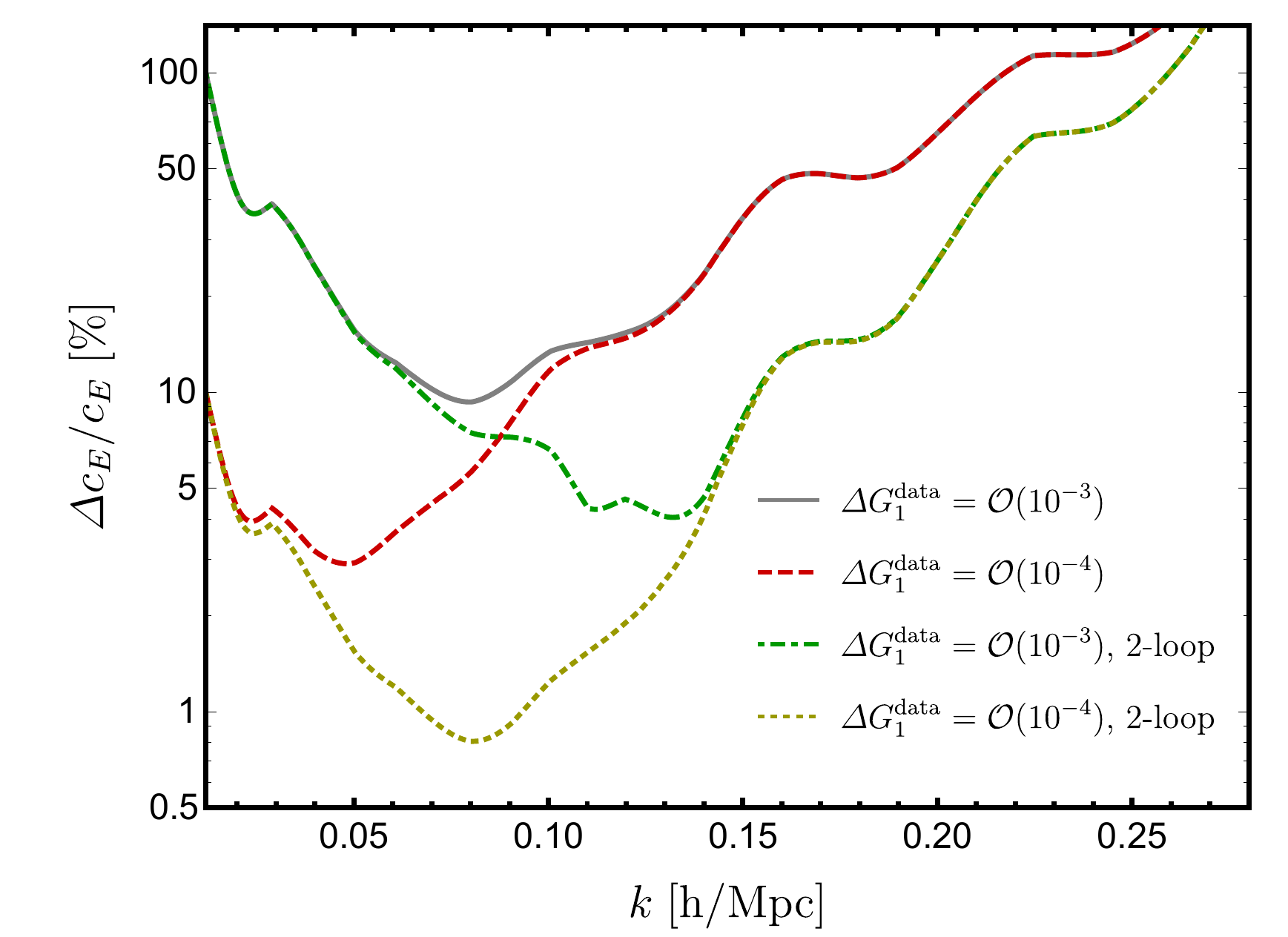}
\caption{Top: coefficient estimator as a function of $k$ with uncertainties from data and theory combined in quadrature (gray) or with uncertainties from data only (black). The blue rectangle denotes our best fit value for $c_E$, with its height and width indicating the uncertainty and the range of $k$ values included in the fit, respectively. Bottom: estimated $\Delta c_E /c_E$ for a measurement at the corresponding $k$ value, assuming $c_E$ is an $\order(k_{\rm NL}^{-2})$ constant. The gray solid line assumes the same uncertainties as in our analysis. The red dashed line assumes an order of magnitude smaller uncertainties for the data, while the green dot-dashed line assumes that two-loop SPT and EFT corrections are known. The yellow dotted curve assumes both these improvements. We use $k_{\rm NL} = 0.34$ h/Mpc for both plots.}
\label{fig:fit}
\end{center}
\end{figure*}

The estimator and its uncertainties are shown in the upper panel of Fig.~\ref{fig:fit}. We include the theory error in \Eq{eq:dGtheory}, and thus the fit can be extended up to $k_\text{NL}$, which we vary to minimize $(\chi^2/{\rm DOF} -1)$. Furthermore, we only include data for $k \geq k_{\rm min} = 0.05$ h/Mpc, since below this value the data seem to have large scatter and potentially underestimated systematic errors (the lowest data point is off-scale in Fig.~\ref{fig:fit}). 
The fit yields $c_E = (-2.11 \pm 0.34)\, \text{Mpc}^2/\text{h}^2$ with $k_{\rm NL} = 0.34$ h/Mpc and $\chi^2/{\rm DOF} = 1.03$.
We find consistent results when varying $k_{\rm min}$ up to $0.08$ h/Mpc, but note that fits with $k_{\rm min} > 0.08$ h/Mpc have poor $\chi^2/{\rm DOF}$. From the EFT predicition for $G_1(k)$ shown in Fig.~\ref{fig:G1}, we can see that the fit is driven by data in the range $k = 0.05 - 0.10$ h/Mpc, and that the data at higher $k$ are consistent within uncertainties.

Our measurement of $c_E$ corresponds to the combination of EFT coefficients in Eq.~(\ref{eq:G1eft}), and we may translate it to a measurement of $\bar{c}_b$. Assuming the value $\bar{c}_s = (9 \pm 0.9)\,\text{Mpc}^2/\text{h}^2$ from Refs.~\cite{Baldauf:2015aha,Baldauf:2015zga}, we find $\bar{c}_b = (-5.03 \pm 1.14)\,\text{Mpc}^2/\text{h}^2$. The bispectrum EFT parameters have been measured in Ref.~\cite{2015JCAP...05..007B}, and their results for
$\bar{c}_{s,1,2,3}$ 
give $c_E = (-3.77 \pm 3.55)\,\text{Mpc}^2/\text{h}^2$ and $\bar{c}_b = (-9.54 \pm 11.73)\,\text{Mpc}^2/\text{h}^2$.
These are consistent with our results, but have large relative uncertainties due to cancellations among $\bar{c}_{1,2,3}$. 

As discussed at the end of the previous section, we may perform a consistency check by assuming the VKPR ansatz and measuring $\bar{c}_s$ (see Appendix~\ref{app:VKPR}). While the difference between the full results in Fig.~\ref{fig:G1} is a few percent, the difference between the one-loop corrections are significantly larger. Performing the same analysis, we find $c_E = (-3.52 \pm 0.31)\,\text{Mpc}^2/\text{h}^2$ with $k_{\rm NL} = 0.36$ h/Mpc and $\chi^2/{\rm DOF} = 1.06$. This translates to $\bar{c}_s = (12.8 \pm 1.1)\,\text{Mpc}^2/\text{h}^2$, which is consistent with previous measurements (e.g., see Refs.~\cite{Baldauf:2015aha,Baldauf:2015zga}).

We have argued for the use of response functions to measure EFT coefficients reliably and precisely.
Let us briefly comment on the precision of the fit presented in this section, and on potential ways to improve it.
The bottom panel in Fig.~\ref{fig:fit} shows estimates for $\Delta c_E/c_E$ for a measurement at the corresponding $k$ value. All curves assume the uncertainties in \Eq{eq:E}, but have different values for $\Delta G_1^{\rm data}$ and $\Delta G_1^{\rm theory}$ as discussed in the following. We have assumed that $c_E$ is an $\mathcal{O}(k_{\rm NL}^{-2})$ constant and $k_{\rm NL} = 0.34 $ h/Mpc.

At large $k$ the theory error dominates, as missing higher-order terms become increasingly important. On the other hand, at low $k$, the error increases as  $1/k^2$, assuming $\Delta G_1^\text{data}$ is constant. Thus, there is an optimal region of $k$ for the measurement of the EFT parameter. For the uncertainties used in our analysis (gray solid curve in the bottom panel of \Fig{fig:fit}), the optimal region is $k= 0.05 - 0.10$ h/Mpc, and gives $\Delta c_E/c_E = 10 - 15 \%$, which is the level of precision we get in our fit.
More precise data for $G_1(k)$ could allow for a better determination of the EFT coefficient. Assuming an order of magnitude improvement in $\Delta G_1^\text{data}$ (red dashed curve in the bottom panel of \Fig{fig:fit}), the optimal region is $k=0.03 - 0.07$ h/Mpc, and gives $\Delta c_E/c_E = 3 - 5 \%$.
Alternatively, including higher-order corrections would reduce the theory uncertainty, but also introduce additional EFT coefficients
(see, e.g., Refs.~\cite{Baldauf:2015aha,Foreman:2015lca}). Assuming that these are known (green dot-dashed curve in the bottom panel of \Fig{fig:fit}), the optimal region is $k = 0.10 - 0.15$ h/Mpc, and gives $\Delta c_E/c_E = 4 - 8 \%$. If both of these improvements are assumed (yellow dotted curve in the bottom panel of \Fig{fig:fit}), the optimal region is $k=0.05 - 0.10$ h/Mpc, and gives $\Delta c_E/c_E = 1 - 2 \%$.
Note that these prospects are optimistic since systematic uncertainties in N-body simulation data are especially difficult to control at low $k$, and higher-order EFT coefficients are not known without resorting to additional measurements or ansatze. Nonetheless, it is interesting that such a generic estimate using only $\Delta c_E$ in Eq.~(\ref{eq:E}) can be made to inform both the optimal region for the measurement and the precision that can be expected. Finally, we comment that, in the context of specifying the full set of EFT contributions for the angle-averaged bispectrum, once the principal shape is determined with less than $\order(10 \%)$ precision, the subdominant shapes become relevant and would also have to be measured.

\subsection{Squeezed Covariance and Anisotropic Responses}
The analysis in the preceding sections employed the relation in \Eq{eq:Rn}, applicable for responses to isotropic background fields. For example, taking $N=4$ for the trispectrum, the relation involves two background modes with wavevectors $\vec{q}_1$ and $\vec{q}_2$, whose directions are separately averaged over.
On the other hand, we are interested in the non-Gaussian covariance of the matter power spectrum which depends on a particular configuration of the tripsectrum given by
\begin{align}\label{eq:cov1}
C_\text{NG}(k,q)=\frac{1}{V}\langle T(\vec{k},-\vec{k},\vec{q},-\vec{q})\rangle,
\end{align}
where $V$ is the volume of the survey, and the average denoted by $\langle \ \rangle$ is over the angle between $\vec{q}$ and $\vec{k}$. 
Hence, to relate the squeezed limit ($q\ll k$) of this observable to a response function, we need to generalize \Eq{eq:Rn} to account for the condition $\vec{q}_2 = - \vec{q}_1 $. To this end, let us consider power spectrum responses to anisotropic long-wavelength backgrounds.
 
We start by generalizing \Eq{eq:R1} for the bispectrum. We define a generalized response
\begin{align}
\mathcal{R}_1(k,\mu)\equiv\lim_{q\to 0}\,\frac{B(\vec{k},\vec{k}',\vec{q})}{P(k)P_L(q)},
\end{align}
where $\mu$ is the cosine of the angle between the $\vec{q}$ and $\vec{k}$ directions, and $\mathcal{R}_1(k,\mu)$ is a response of the power spectrum to an anisotropic long-wavelength background. In the presence of a long-wavelength perturbation $\delta_{\vec{q}}(\vec{x},t)\sim \delta_L(t) e^{i\vec{q}\cdot\vec{x}}$ with wavevector $\vec{q}$, the power spectrum will also be anisotropic and, at leading order in $q/k$, can be written as 
\begin{align}
P\left(\vec{k}|\delta_\vec{q}\right)=P_\text{iso}(k|\delta_L)+A_1(k)L_2(\mu)\delta_L,
\end{align}
where  $L_2(\mu)=(3\mu^2-1)/2$ is a Legendre polynomial and the time-dependence is implicit. Then, $\mathcal{R}_1(k,\mu)$ is the linear response of $P\left(\vec{k}|\delta_\vec{q}\right)$ to a change in the background amplitude:
\begin{align}\label{eq:RR1}
\mathcal{R}_1(k,\mu)=\frac{1}{P(k)}\frac{\df P\left(\vec{k}|\delta_\vec{q}\right)}{\df\delta_L}\Bigg|_{\delta_L=0}=R_1(k)+A_1(k)L_2(\mu),
\end{align}
where $R_1(k)= 1/P(k)\, \df P_\text{iso}(k|\delta_L)/\df\delta_L|_{\delta_L=0}$ is the isotropic response defined in \Eq{eq:R1}, and $A_1(k)$ is the anisotropic response in the presence of a single directional background mode. 

Let us now consider a similar generalization for the trispectrum. We define a response
\begin{align}\label{eq:R2}
\mathcal{R}_2(k,\mu_1,\mu_2,\mu_{12})\equiv\lim_{q_{i} \to 0}\frac{T(\vec{k},\vec{k}',\vec{q}_1,\vec{q}_2)}{P(k)P_L(q_1)P_L(q_2)},
\end{align}
where $\mu_{i}$ is the cosine of the angle between $\vec{k}$ and $\vec{q}_{i}$, and $\mu_{12}$ is the cosine of the angle between $\vec{q}_1$ and $\vec{q}_2$. 
While the limit in the isotropic case is independent of the ratio $q_1/q_2$, the limit in the anisotropic case is not, and we take $q_1/q_2=1$. The power spectrum in the presence of two directional soft modes $\delta_{\vec{q}_1}(\vec{x},t)\sim \delta_L(t) e^{i\vec{q}_1\cdot\vec{x}}$ and $\delta_{\vec{q}_2}(\vec{x},t)\sim \delta_L(t) e^{i\vec{q}_2\cdot\vec{x}}$, having the same amplitude $\delta_L$ and wavevectors $\vec{q}_1$ and $\vec{q}_2$ with $q_1 = q_2$, can be written, at leading order in $q_{1,2}/k$, as
\begin{align}
P\left(\vec{k}|\delta_{\vec{q}_1}\delta_{\vec{q}_2}\right)=P_\text{iso}(k|\delta_L)+\frac{1}{2}\sum_i A_2^i(k)\,\Theta_i(\mu_1,\mu_2,\mu_{12})\delta_L^2\,,
\end{align}
where the functions $\Theta_i(\mu_1,\mu_2,\mu_{12})$ form an angular basis, and
vanish upon averaging over $\mu_{1}$, $\mu_2$ and $\mu_{12}$.
The relation for the response is then given by
\begin{align}\label{eq:RR2}
\mathcal{R}_2(k,\mu_1,\mu_2,\mu_{12})=\frac{1}{P(k)}\frac{\df^2 P\left(\vec{k}|\delta_{\vec{q}_1}\delta_{\vec{q}_2}\right)}{\df\delta_L^2}\Bigg|_{\delta_L=0}=R_2(k)+\sum_i A_2^i(k)\,\Theta_i(\mu_1,\mu_2,\mu_{12}),
\end{align}
where $R_2(k)= 1/P(k)\,\df^2 P_\text{iso}(k|\delta_L)/\df\delta_L^2|_{\delta_L=0}$ is the isotropic response, and the functions $A_2^i(k)$ are the anisotropic responses in the presence of two directional background modes. A proof for Eqs.~(\ref{eq:RR1}) and~(\ref{eq:RR2}) can be constructed along the lines of the proof for the isotropic case given in Appendix~A of Ref.~\cite{Wagner:2015gva}.

For the case of the covariance, we fix the relative orientation of the two background modes, $\vec{q}_2=-\vec{q}_1$, and average over the remaining angle $\mu_1$. The response relevant for the squeezed covariance is thus defined, through Eqs.~(\ref{eq:R2}) and~(\ref{eq:RR2}), as
\begin{align}\label{eq:C}
\mathcal{C}(k) &\equiv \lim_{q \to 0}\frac{ V C_\text{NG}(k,q)}{P(k)P_L(q)^2}
=R_2(k)+\sum_i A_2^i(k)\,\langle\Theta_i(\mu_1,-\mu_1,-1)\rangle \, ,
\end{align}
where, in contrast to the isotropic case, the average of the functions $\Theta_i$ no longer vanishes.

The measurement, e.g.,~from N-body simulations, of the responses $R_2(k)$ and $A_2^i(k)$ would determine $\mathcal{R}_2(k,\mu_1,\mu_2,\mu_{12})$ and thus $\mathcal{C}(k)$ as a particular case. 
The anisotropic responses can be measured by considering two background modes with fixed directions and a common amplitude, and then projecting the second-order response of the power spectrum to a change in the amplitude onto the angular basis in \Eq{eq:RR2}. 
They can also be measured as first-order responses of the power spectrum 
to a change in the directions of the tidal fields, keeping the amplitude fixed.
Note that in the presence of directional background modes, the standard separate-universe approach does not apply, but can be generalized by considering an anisotropic Bianchi cosmology (see, e.g., Refs.~\cite{Dai:2015rda,Garny:2015oya}).

Alternatively, it may be possible to measure $\mathcal{C}(k)$ without first determining the anisotropic responses, by considering a composite background configuration of two density perturbations: $\eta(\vec{q}) \sim \delta_\vec{q}\delta_{-\vec{q}}$.
Then, the function $\mathcal{C}(k)$ would be given by the first-order isotropic response of the power spectrum $P\left(k| \eta_\vec{q} \right)$ to a change in the amplitude of $\eta_\vec{q}$, after averaging over the single direction $\vec{q}$.

Let us now collect perturbative results for $\mathcal{R}_2(k)$ and $\mathcal{C}(k)$. 
The tree-level result is
\begin{align*}
R_2^\text{tree}(k)&=\frac{8420}{1323}-\frac{100k}{63}\frac{P_L'(k)}{P_L(k)}+\frac{k^2}{9}\frac{P_L''(k)}{P_L(k)} \,, &
A_1^\text{tree}(k)&=\frac{352}{147}-\frac{29k}{14}\frac{P_L'(k)}{P_L(k)} \,, \\
A_2^\text{tree}(k)&=\frac{8}{21}-\frac{3k}{7}\frac{P_L'(k)}{P_L(k)} \,, &
A_3^\text{tree}(k)&=\frac{26}{7}\,, \\
A_4^\text{tree}(k)&=\frac{656}{147}-\frac{23k}{7}\frac{P_L'(k)}{P_L(k)}+k^2\frac{P_L''(k)}{P_L(k)}\,, &
A_5^\text{tree}(k)&=\frac{16}{21}\,, \\
A_6^\text{tree}(k)&=-\frac{16}{7}+\frac{10k}{7}\frac{P_L'(k)}{P_L(k)}\,, &
A_7^\text{tree}(k)&=\frac{20}{21}-\frac{11k}{14}\frac{P_L'(k)}{P_L(k)} \,,\numberthis
\end{align*} 
where the isotropic response $R_2(k)$ had been calculated at tree level in Ref.~\cite{Wagner:2015gva}, and the angular functions are given by 
\begin{align*}
\Theta_1(\mu_1,\mu_2,\mu_{12})&=\mu_1^2+\mu_2^2-2/3\,, &
\Theta_2(\mu_1,\mu_2,\mu_{12})&=\mu_1\mu_2\,, \\
\Theta_3(\mu_1,\mu_2,\mu_{12})&=\mu_{12},&
\Theta_4(\mu_1,\mu_2,\mu_{12})&=\mu_1^2\mu_2^2-1/9,\\
\Theta_5(\mu_1,\mu_2,\mu_{12})&=\mu_{12}^2-1/3,&
\Theta_6(\mu_1,\mu_2,\mu_{12})&=\mu_{12}\mu_1\mu_2-1/9,\\
\Theta_7(\mu_1,\mu_2,\mu_{12})&=\mu_{12}(\mu_1^2+\mu_2^2).\numberthis
\end{align*}
The covariance response follows upon fixing the relative angle between the background modes, and averaging over the remaining angle:
\begin{align}
\mathcal{C}^\text{tree}(k)=\frac{5038}{2205}-\frac{94k}{105}\frac{P_L'(k)}{P_L(k)}+\frac{k^2}{5}\frac{P_L''(k)}{P_L(k)}.
\end{align}

The one-loop correction to $\mathcal{C}(k)$ can be obtained by taking the squeezed limit of the one-loop correction to the covariance calculated in Ref.~\cite{2015arXiv151207630B}. We do not show the SPT corrections here, but we focus on the corresponding EFT contributions to connect with the discussion in Sec.~\ref{subsec:alignment} on the EFT coefficients in the squeezed limit. For the covariance, we need the stress tensor through NNLO in fields, given in \Eq{eq:stresstensor}. As derived in Ref.~\cite{2015arXiv151207630B}, assuming that the speed of sound and the three bispectrum counterterms are known from lower orders, there are three new independent operators, with coefficients $\bar{c}_{4,5,6}$ (see Eq.~(24) of Ref.~\cite{2015arXiv151207630B} or \App{app:EFT} for the definition of the corresponding operators). In the squeezed limit, these three operators become degenerate, and the EFT counterterm is given by $\mathcal{C}^\text{EFT}(k)=\mathcal{C}^\text{EFT}_\text{lower}(k)+\mathcal{C}^\text{EFT}_\text{NNLO}(k)$, with
\vspace{0.1cm}
\begin{align*}
\mathcal{C}^\text{EFT}_\text{lower}(k)&=\bar{c}_sk^2\left(-\frac{61858}{257985}+\frac{388k}{3465}\frac{P_L'(k)}{P_L(k)}-\frac{2k^2}{45}\frac{P_L''(k)}{P_L(k)}\right) +\bar{c}_1k^2\left(-\frac{208}{693}+\frac{16k}{99}\frac{P_L'(k)}{P_L(k)}\right)\\
&\quad +\bar{c}_2k^2\left(-\frac{5216}{10395}+\frac{64k}{495}\frac{P_L'(k)}{P_L(k)}\right)+\bar{c}_3k^2\left(-\frac{3632}{10395}+\frac{16k}{165}\frac{P_L'(k)}{P_L(k)}\right) \,, \\[3pt]
\mathcal{C}^\text{EFT}_\text{NNLO}(k)&=\frac{2\sqrt{101785}}{1365}\underbrace{\frac{48\bar{c}_4+16\bar{c}_5-315\bar{c}_6}{\sqrt{101785}}}_{\bar{c}_t}k^2 \,, \numberthis
\end{align*}
where $\mathcal{C}^\text{EFT}_\text{lower}(k)$ contains the contribution from the propagation of the power spectrum and bispectrum counterterms, and $\bar{c}_t$ is the normalized combination of $\bar{c}_{4,5,6}$ in Eq.~(\ref{eq:align}).
Assuming that $\bar{c}_s$ and $\bar{c}_{1,2,3}$ are known, and if data for the relevant responses become available, e.g., $A_2^i(k)$ in Eq.~(\ref{eq:RR2}), then we can measure $\bar{c}_t$. As discussed in \Sec{subsec:alignment}, $\bar{c}_t$ determines the principal EFT shape for the covariance over the entire kinematic range, even outside the squeezed region.

\section{Conclusions}
\label{sec:conclusion}
We have explored two strategies for developing the EFT of LSS as a viable framework for meaningful comparison between theory and data. The first strategy is to identify principal EFT shapes, i.e.~shapes that dominate over the relevant kinematic domain, by accounting for correlations among EFT contributions. The second strategy is to employ response functions for measuring EFT coefficients. We focused on an interesting connection between the two strategies: measuring principal EFT coefficients from response functions. 

In particular, we showed that, for the angle-averaged bispectrum and the non-Gaussian covariance of the matter power spectrum, the EFT contributions are hierarchical (Fig.~\ref{fig:bishapes}), and that the principal EFT coefficient is well approximated by the combination of coefficients in the squeezed limit (Eq.~(\ref{eq:align})). For the angle-averaged bispectrum, we then measured the principal coefficient $\bar{c}_b$ from N-body simulation data for the growth-only response (Fig.~\ref{fig:G1}). For the covariance, we confirmed that the principal coefficient $\bar{c}_t$, obtained from our orthogonalization procedure, agrees with the result from applying a PCA directly on data. We also provided the set of anisotropic responses that can be used to measure $\bar{c}_t$ (\Eq{eq:C}). 

Moving forward, it would be interesting to check the accuracy of our method by comparing the one-loop prediction for the angle-averaged bispectrum, obtained with the principal EFT shape measured here, to bispectrum N-body data away from the squeezed configuration. Our measurement of $\bar{c}_b$ fixes one combination of the bispectrum EFT parameters, and this could be used to improve measurements of the remaining parameters by eliminating degenerate fit solutions (in Refs.~\cite{2014arXiv1406.4143A,2015JCAP...05..007B}, fits of $\bar{c}_{1,2,3}$ yielded multiple solutions). Additionally, the same strategy applied here for the angle-averaged bispectrum, could also be applied for the angle-averaged trispectrum, and the corresponding EFT shape could be measured from the second-order growth-only response, $G_2(k)$, for which N-body data are available from Ref.~\cite{Wagner:2015gva}.

There are also a number of directions to pursue for developing the two strategies presented here as independent tools. For example, it would be interesting to extend the orthogonalization analysis to other observables and to incorporate uncertainties and restictions on the kinematic domain relevant for actual data surveys, such as for the detection of primordial non-Gaussianities~\cite{Welling:2016dng}. On the other hand, aside from improving the precision and variety of responses determined from N-body simulations, we could map out the combinations of EFT coefficients that can be accessed from responses of low-order correlation functions, and investigate the factorization of amplitudes for LSS observables in the squeezed limit.

\section*{Acknowledgements}
We thank Chi-Ting Chiang for sharing simulation data and for correspondence regarding Ref.~\cite{Wagner:2015gva}. We also thank Simone Ferraro and Enrico Pajer for helpful discussions and comments on the manuscript. DB and MS are supported under contract DE-AC02-05CH11231 and by WPI, MEXT, Japan.

\appendix
\section{SPT and EFT Essentials}
\label{app:EFT}
We briefly review the SPT and EFT of LSS frameworks, and collect results relevant for the discussion in this paper. We refer the reader to Ref.~\cite{2002PhR...367....1B} for a more detailed review on SPT, and to Refs.~\cite{2012JCAP...07..051B,2014PhRvD..89d3521H,2012JHEP...09..082C} for the derivation of the EFT of LSS framework.

Starting from the collisionless Boltzmann equation, and defining long-wavelength density and velocity fields in terms of smoothed momenta of the probability distribution, one derives the standard hydrodynamic equations which describe dark matter on large scales:
\begin{align}
\partial_\tau \delta(\vec{k}) + \theta(\vec{k}) &= S_\alpha,\label{deltaeqn2}\\
 \partial_\tau \theta(\vec{k}) +  \mathcal{H} \theta(\vec{k}) + \frac{3}{2} \mathcal{H}^2\Omega_m \delta(\vec{k}) &= S_\beta -\partial_i \left(\frac{\partial_j \uptau^{ij}}{1+\delta}\right), \label{thetaeqn2} \\
\partial_\tau \omega^i(\vec{k}) +  \mathcal{H} \omega^i(\vec{k})&= S_\gamma -\epsilon^{ijk} \partial_j \left(\frac{\partial_b \uptau^{kb}}{1+\delta}\right),\numberthis\label{omegaeq2}
\end{align}
where $\delta(\vec{k})$ is the smoothed density perurbation, $\theta(\vec{k})\equiv\partial_iv^i(\vec{k})$ and $\omega^i(\vec{k})=\epsilon^{ijr}\partial_jv_r(\vec{k})$ are the velocity divergence and vorticity respectively, and for simplicity we have suppressed the time dependence. Derivatives are taken with respect to conformal time, and $\mathcal{H}$ is the conformal Hubble parameter. Note that Eqs.~(\ref{deltaeqn2}-\ref{omegaeq2}) are valid so long as one calculates correlators of $\delta$ only. As discussed, e.g., in Ref.~\cite{Bertolini:2016bmt}, correlators involving $\theta$ require additional counterterms. The standard SPT mode-coupling functions, $S_\alpha$, $S_\beta$ and $S_\gamma$, are given by
\begin{align*}
S_\alpha&=- \int d^3q \Big( \alpha (\vec{q}, \vec{k}-\vec{q}) \theta(\vec{q})  - \alpha_\omega^i (\vec{q}, \vec{k}-\vec{q})\omega_i(\vec{q}) \Big) \delta(\vec{k}- \vec{q}),\numberthis\\
S_\beta&=-\int d^3 q \bigg( \beta(\vec{q}, \vec{k}-\vec{q}) \theta(\vec{q}) \theta(\vec{k} - \vec{q})
-\vec{\beta}_\omega^i(\vec{q}, \vec{k}-\vec{q})  \vec{\omega}_i(\vec{q})\theta(\vec{k} - \vec{q})\\
&\quad+ \beta_{\omega\omega}^{ij}(\vec{q}, \vec{k}-\vec{q}) \,\omega_i (\vec{q})\,\omega_j( \vec{k}-\vec{q})\bigg),\numberthis\\
S_\gamma&=-\int d^3 q  \bigg(-\gamma^{ij}_\omega(\vec{q}, \vec{k}-\vec{q}) \omega_j(\vec{q}) \theta(\vec{k}-\vec{q})
+\gamma^{ijk}_{\omega\omega}(\vec{q}, \vec{k}-\vec{q}) \omega_j(\vec{q}) \omega_k(\vec{k}-\vec{q}) \bigg),\numberthis
\end{align*}
with the kernels
\begin{align*}\label{eq:couplings}
\alpha(\vec{k}_1, \vec{k}_2) &= \frac{\vec{k}_1 \cdot \vec{k}}{k_1^2},
&\alpha_\omega^i(\vec{k}_1, \vec{k}_2)& = \frac{(\vec{k}_2 \times \vec{k}_1)^i}{k_1^2},\\
 \beta(\vec{k}_1, \vec{k}_2)& = \frac{k^2 (\vec{k}_1\cdot \vec{k}_2)}{2\, k_1^2 k_2^2},
&\beta_\omega^i(\vec{k}_1, \vec{k}_2)& = \frac{(2 (\vec{k}_1\cdot\vec{k}_2) + k_2^2)(\vec{k}_2\times \vec{k}_1)^i}{k_1^2 k_2^2},\\
 \beta_{\omega\omega}^{ij}(\vec{k}_1, \vec{k}_2) &= \frac{(\vec{k}_2\times \vec{k}_1)^i (\vec{k}_1\times \vec{k}_2)^j}{k_1^2 k_2^2},
&\gamma^{ij}_\omega (\vec{k}_1, \vec{k}_2) & =\frac{k_2^ik^j-(\vec{k}\cdot\vec{k}_2)\delta^{ij}}{k_2^2},\\
\gamma_{\omega \omega}^{ijk}(\vec{k}_1, \vec{k}_2) &=  { \epsilon^{imj}k_m k_1^k-(\vec{k} \times \vec{k}_1)^i \delta^{jk} \over k_1^2}\numberthis \,,
\end{align*}
where $\vec{k} = \vec{k}_1 + \vec{k}_2$. The stress tensor $\uptau^{ij}$ includes the EFT counterterms, and can be expanded in powers of fields and derivatives. Here, we are interested in the stress tensor at leading order in the derivative expansion, including up to three-field terms (i.e.~up to trispectrum counterterms).
There is one independent operator at one-field order (LO), three new independent operators at two-field order (NLO), and eight new independent operators at three-field order (NNLO). Following the parametrization of Refs.~\cite{2015arXiv151207630B,Bertolini:2016bmt}, we get
\begin{align*}\label{eq:stresstensor}
&k_i { \uptau}^{ij}  = \, \bar{c}_s^\delta k^j  \delta({\vec k})
+ \int \df \vec{q} \sum_{n=1}^3 \bar{c}_n^{\delta\delta}  \delta({{\vec q}})\delta(\vec{k} - {{\vec q}}) k_i e^{ij}_n (\vec{q}, \vec{k} - \vec{q})\\
&\quad\quad\quad + \int \df \vec{q} \sum_{n=2}^3 {\bar{c}_n^{\theta\theta}\over \mathcal{H}^2f^2}  \theta({{\vec q}})\theta(\vec{k} - {{\vec q}})) k_i e^{ij}_n (\vec{q}, \vec{k} - \vec{q})\\
&\quad\quad\quad  +  \int \df \vec{q}_1 \df \vec{q}_2  \sum_{n=1}^{6} \bar{c}_n^{\delta\delta\delta}   \delta({{\vec q}_1}) \delta({{\vec q}_2}) \delta(\vec{k} - {{\vec q}_1} - \vec{q}_2)  k_i E^{ij}_n (\vec{q}_1, \vec{q}_2, \vec{k} - {{\vec q}_1} - \vec{q}_2) \,. \numberthis 
\end{align*} 
The functions $e_n^{ij}$ and $E_n^{ij}$ are given by
\begin{align*}\label{eq:operators}
E_1^{ij} (\vec{q}_1, \vec{q}_2, \vec{q}_3)&= e_1^{ij} (\vec{q}_1, \vec{q}_2) =\delta^{ij} \,, &
E_2^{ij} (\vec{q}_1, \vec{q}_2, \vec{q}_3)&= e_2^{ij} (\vec{q}_1, \vec{q}_2) = {q_1^i q_1^j \over q_1^2} \,, \\
E_3^{ij} (\vec{q}_1, \vec{q}_2, \vec{q}_3)&= e_3^{ij} (\vec{q}_1, \vec{q}_2) = {q_1^{\{ i} q_2^{j\}} q_1^a q_2^a \over q_1^2 q_2^2} \,, &
E_4^{ij} (\vec{q}_1, \vec{q}_2, \vec{q}_3) & = {\delta^{ij}  (q_1^a q_2^a)^2 \over q_1^2 q_2^2} \,, \\
E_5^{ij} (\vec{q}_1, \vec{q}_2, \vec{q}_3)&= {q_1^i q_1^j (q_2^a q_3^a)^2 \over q_1^2 q_2^2 q_3^2 }  \,, &
E_6^{ij} (\vec{q}_1, \vec{q}_2, \vec{q}_3)&= {q_1^{\{ i} q_2^{j\}} q_1^a q_3^a q_2^b q_3^b \over q_1^2 q_2^2 q_3^2 }.\, \numberthis
\end{align*}
For each operator in \Eq{eq:stresstensor} we have introduced a coefficient $\bar{c}$ with dimensions $[k]^{-2}$
and time dependence that matches the SPT loops, $\bar{c}=[\mathcal{H}f(\tau)D(\tau)]^2c$, where c
is time independent. In the above equation $D(\tau)$ is the linear growth function, and $f(\tau)=1/\mathcal{H}\,\df\log D(\tau)/\df\tau$. The equations of motion can be solved in perturbation theory using the EdS-like anstatz
for the growing modes:
\begin{align}
\delta(\vec{k},\tau)&=\sum_{n=1}^\infty\left( D^n(\tau)\,\delta_n(\vec{k})+\varepsilon\,D^{n+2}(\tau)\,\tilde{\delta}_n(\vec{k})\right),\label{eq:ansatzd}\\
\theta(\vec{k},\tau)&=-\mathcal{H}f(\tau)\sum_{n=1}^\infty\left( D^n(\tau)\,\theta_n(\vec{k})+\varepsilon\,D^{n+2}(\tau)\,\tilde{\theta}_n(\vec{k})\right),\label{eq:ansatzt}\\
\omega^i(\vec{k},\tau)&=-\mathcal{H}f(\tau)\sum_{n=2}^\infty \varepsilon\,D^{n+2}(\tau)\,\tilde{\omega}_{n}^i(\vec{k}),\label{eq:ansatzw}
\end{align}
where $\varepsilon$ tracks the EFT corrections. At each order, the fields can be expanded in powers of the linear density perturbation according to
\begin{align*}
\begin{pmatrix} 
\delta_n(\vec{k}) \\ 
\theta_n(\vec{k}) \\
\tilde{\delta}_n(\vec{k}) \\
\tilde{\theta}_n(\vec{k})\\
\tilde{\omega}_n^i(\vec{k})  
\end{pmatrix}& = 
\int \dbar^{\,3}q_1 ... \ \dbar^{\,3}q_n \, \begin{pmatrix} F_n(\vec{q}_1,...,\vec{q}_n) \\ G_n(\vec{q}_1,...,\vec{q}_n) \\ \widetilde{F}_n(\vec{q}_1,...,\vec{q}_n) \\
\widetilde{G}_n(\vec{q}_1,...,\vec{q}_n)\\
\widetilde{G}_{n}^{\omega i}(\vec{q}_1,...,\vec{q}_n)
\end{pmatrix}
(2 \pi)^3 \delta_D\left(\vec{k}-\sum_{i=1}^n{\vec{q}_i}\right) \delta_1(\vec{q}_1)... \delta_1(\vec{q}_n),
 \numberthis \label{eq:kernels}\end{align*} 
where $\dbar^{\,3}q\equiv \df \vec{q} /(2\pi)^3$.
The SPT kernels $F_n$ and $G_n$ can be found in Ref.~\cite{1986ApJ...311....6G}, and the EFT kernels up to NNLO are listed below. At LO we find
\begin{align} 
\widetilde{F}_1(\vec{k}) = -\frac{1}{9} c_s k^2,\quad\quad
\widetilde{G}_1(\vec{k}) = -\frac{1}{3}c_s k^2, 
\end{align}
where, for simplicity of notation, we have renamed $c_s^\delta=c_s$. At NLO we find

\begin{align*} \widetilde{F}_2(\vec{k}_1, \vec{k}_2) &= \,\frac{3}{11} \alpha (\vec{k}_1, \vec{k}_2) \left( \widetilde{G}_1 (\vec{k}_1) + \tilde{F}_1(\vec{k}_2)\right) + \frac{2}{33} \beta (\vec{k}_1, \vec{k}_2) \left(\widetilde{G}_1 (\vec{k}_1) + \widetilde{G}_1 (\vec{k}_2)\right)\\
&\quad -\frac{2}{33} \,c_s\left( k^2 F_2(\vec{k}_1, \vec{k}_2) -\vec{k}\cdot\vec{k}_2\right)- \frac{2}{33} \sum_{n=1}^3c_n\, k_ik_j\,e_n^{ij} (\vec{k}_1, \vec{k}_2),\\
\numberthis\label{eq:f2tilde} \\ 
\widetilde{G}_2(\vec{k}_1, \vec{k}_2)  &= \,\frac{1}{11} \alpha (\vec{k}_1, \vec{k}_2) \left( \widetilde{G}_1 (\vec{k}_1) + \widetilde{F}_1(\vec{k}_2)\right) + \frac{8}{33} \beta (\vec{k}_1, \vec{k}_2) \left(\widetilde{G}_1 (\vec{k}_1) + \widetilde{G}_1 (\vec{k}_2)\right)\\
&\quad -\frac{8}{33} \,c_s\left( k^2 F_2(\vec{k}_1, \vec{k}_2) -\vec{k}\cdot\vec{k}_2\right) - \frac{8}{33} \sum_{n=1}^3c_n\, k_ik_j\,e_n^{ij} (\vec{k}_1, \vec{k}_2),  
\numberthis\label{eq:g2tilde} \end{align*}
where $\vec{k}=\vec{k}_1+\vec{k}_2$, and $c_{1,2,3}$ are defined as in \cite{2015arXiv151207630B} to be
\begin{align}
\label{eq:c123}
c_1 \equiv c_1^{\delta\delta}, \quad\quad
c_2 \equiv c_2^{\delta\delta}+c_2^{\theta\theta}, \quad\quad
c_3 \equiv c_3^{\delta\delta}+c_3^{\theta\theta}. \quad\quad
\end{align} 
Finally, at NNLO we find
\begin{align*}
\widetilde{F}_3(\vec{k}_1,\vec{k}_2,\vec{k}_3) &=  \, \frac{11}{52} \alpha( \vec{k}_1,  \vec{k}_2+ \vec{k}_3) \left[ \widetilde{G}_1 ( \vec{k}_1) F_2( \vec{k}_2,  \vec{k}_3) + \widetilde{F}_2 ( \vec{k}_2,  \vec{k}_3) \right] + \frac{11}{52} \alpha( \vec{k}_1+ \vec{k}_2,  \vec{k}_3) \Big[ \widetilde{G}_2( \vec{k}_1,  \vec{k}_2)\\
&\quad+G_2( \vec{k}_1,  \vec{k}_2) \widetilde{F}_1(\vec{k}_3)\Big] + \frac{1}{26} \beta( \vec{k}_1,  \vec{k}_2 +  \vec{k}_3) \left[\widetilde{G}_1 ( \vec{k}_1) G_2( \vec{k}_2,  \vec{k}_3) + \widetilde{G}_2 ( \vec{k}_2,  \vec{k}_3) \right]\\
&\quad + \frac{1}{26} \beta( \vec{k}_1+  \vec{k}_2 ,  \vec{k}_3) \left[ \widetilde{G}_2( \vec{k}_1,  \vec{k}_2) +G_2( \vec{k}_1,  \vec{k}_2) \widetilde{G}_1(\vec{k}_3)\right]\\
&\quad+\frac{1}{26} \beta^i_\omega (\vec{k}_1+\vec{k}_2, \vec{k}_3)\, \widetilde{G}_{2i}^{\omega}(\vec{k}_1, \vec{k}_2)
-\frac{11}{52} \alpha_\omega^i (\vec{k}_1 + \vec{k}_2, \vec{k}_3)\, \widetilde{G}_{2i}^{\omega} (\vec{k}_1, \vec{k}_2)\\
&\quad - \frac{1}{26 } c_s\left( k^2 F_3(\vec{k}_1, \vec{k}_2, \vec{k}_3)-\vec{k} \cdot (\vec{k}_2+\vec{k}_3) F_2(\vec{k}_2, \vec{k}_3)
+(1 - F_2(\vec{k}_1, \vec{k}_2) ) (\vec{k}\cdot\vec{k}_3)\right)\\
&\quad+\frac{1}{26} \sum_{n=1}^3 c_n\, k_i(k_{2}+k_{3})_j\,e_n^{ij} (\vec{k}_2, \vec{k}_3)-\frac{1}{26}  \sum_{n=1}^5c_n\, k_ik_j\,R_n^{ij} (\vec{k}_1,\vec{k}_2, \vec{k}_3)\\ 
&\quad-\frac{1}{26}  \sum_{n=1}^{6} d_n \,k_ik_j E_n^{ij} (\vec{k}_1, \vec{k}_2, \vec{k}_3),\numberthis \label{eq:f3tilde}
\end{align*}
where again $\vec{k}=\vec{k}_1+\vec{k}_2+\vec{k}_3$. The coefficients $c_{1,2,3}$ are defined in Eq.~\eqref{eq:c123}, and $c_{4,5}$ are defined as in Ref.~\cite{2015arXiv151207630B} to be
\begin{align}
\label{eq:c45}
c_4 \equiv c_2^{\theta\theta}+\frac{5}{2}(c_3^{\delta\delta}+c_3^{\theta\theta}), \quad\quad
c_5 \equiv c_3^{\theta\theta}-\frac{5}{2}(c_3^{\delta\delta}+c_3^{\theta\theta}).
\end{align} 
For simplicity of notation we also have renamed $c^{\delta\delta\delta}_{1...6}=d_{1...6}$. 
The kernel $\widetilde{G}_{2i}^{\omega}$ gives the EFT NLO contribution to the vorticity and it is given by
\begin{align*}
&\widetilde{G}_{2i}^{\omega}(\vec{k}_1,\vec{k}_2)= -\frac{2}{9} \epsilon_{ijm} k^j\sum_{n=1}^3 c_n\,k_l\,e_n^{lm}(\vec{k}_1, \vec{k}_2). \numberthis\label{eq:Gomega}
\end{align*}
The functions $R_{1...5}^{ij}(\vec{k}_1,\vec{k}_2,\vec{k}_3)$ are defined as
\begin{align*}
R_1^{ij}(\vec{k}_1,\vec{k}_2,\vec{k}_3)&=F_2(\vec{k}_2,\vec{k}_3) \,e_1^{ij}(\vec{k}_1, \vec{k}_2+ \vec{k}_3)+F_2(\vec{k}_1,\vec{k}_2) \,e_1^{ij}(\vec{k}_1+\vec{k}_2, \vec{k}_3),\\
R_2^{ij}(\vec{k}_1,\vec{k}_2,\vec{k}_3)&=F_2(\vec{k}_2,\vec{k}_3) \,e_2^{ij}(\vec{k}_1, \vec{k}_2+ \vec{k}_3)+F_2(\vec{k}_1,\vec{k}_2) \,e_2^{ij}(\vec{k}_1+\vec{k}_2, \vec{k}_3),\\
R_3^{ij}(\vec{k}_1,\vec{k}_2,\vec{k}_3)&=\frac{5}{2}\left[F_2(\vec{k}_2,\vec{k}_3)-G_2(\vec{k}_2,\vec{k}_3)\right]e_2^{ij}(\vec{k}_1, \vec{k}_2+ \vec{k}_3)\\
&\quad+\frac{5}{2}\left[F_2(\vec{k}_1,\vec{k}_2)-G_2(\vec{k}_1,\vec{k}_2)\right]e_2^{ij}(\vec{k}_1+\vec{k}_2,\vec{k}_3)\\
&\quad-\frac{1}{2}\left[3F_2(\vec{k}_2,\vec{k}_3)-5G_2(\vec{k}_2,\vec{k}_3)\right]e_3^{ij}(\vec{k}_1, \vec{k}_2+ \vec{k}_3)\\
&\quad-\frac{1}{2}\left[3F_2(\vec{k}_1,\vec{k}_2)-5G_2(\vec{k}_1,\vec{k}_2)\right]e_3^{ij}(\vec{k}_1+\vec{k}_2,\vec{k}_3),\\
R_4^{ij}(\vec{k}_1,\vec{k}_2,\vec{k}_3)&=\left[G_2(\vec{k}_2,\vec{k}_3)-F_2(\vec{k}_2,\vec{k}_3)\right]e_2^{ij}(\vec{k}_1,\vec{k}_2+\vec{k}_3)\\
&\quad+\left[G_2(\vec{k}_1,\vec{k}_2)-F_2(\vec{k}_1,\vec{k}_2)\right]e_2^{ij}(\vec{k}_1+\vec{k}_2,\vec{k}_3),\\
R_5^{ij}(\vec{k}_1,\vec{k}_2,\vec{k}_3)&=\left[G_2(\vec{k}_2,\vec{k}_3)-F_2(\vec{k}_2,\vec{k}_3)\right]e_3^{ij}(\vec{k}_1,\vec{k}_2+\vec{k}_3)\\
&\quad+\left[G_2(\vec{k}_1,\vec{k}_2)-F_2(\vec{k}_1,\vec{k}_2)\right]e_3^{ij}(\vec{k}_1+\vec{k}_2,\vec{k}_3).\numberthis
\end{align*}
Note that the kernels listed above are not symmetric in their arguments, and need to be symmetrized when calculating amplitudes.
One can use a convenient diagrammatic representation for deriving the amplitudes that contribute to correlators at any given perturbative order. Details can be found, e.g., in Ref.~\cite{1996ApJS..105...37S}. 

As derived in Refs.~\cite{2015arXiv151207630B,Bertolini:2016bmt}, for the covariance configuration of the trispectrum there are only three NNLO new operators, assuming that $c_s, c_1, c_2$ and $c_3$ are known. Following the remapping of Eq.~(24) of Ref.~\cite{2015arXiv151207630B}, we can set $d_{2...6}=0$ and 
\begin{align}
d_1\to c_6-\frac{2062}{2079}c_1-\frac{14}{1485}c_3.
\end{align}

\section{EFT Shapes for Bispectrum and Covariance}
\label{app:shapes}
The shapes $S_i$ used in the analysis in Sec.~\ref{sec:shapes} are obtained by computing the EFT counterterm diagrams in Eqs.~(\ref{eq:shapesBi}) and~(\ref{eq:shapesCov}). In our diagrams below, we denote SPT and EFT kernels (or vertices) with black circles and gray squares, respectively, and the linear power spectrum (propagator) with dashed lines. For the angle-averaged bispectrum, we have
\begin{align*}
S_1(k_1,k_2) &= - \frac{2}{33} \int_{-1}^1 \text{d} \mu \,   k_a^2 P_L(k_b) P_L(k_c) + \text{2 permutations}  
\,, \\
S_2(k_1,k_2) &= - \frac{1}{132} \int_{-1}^1 \text{d} \mu \,  \frac{ k_a^4 \left(k_b^2+k_c^2\right)-2 k_a^2 \left(k_b^2-k_c^2\right)^2+\left(k_b^2-k_c^2\right)^2 \left(k_b^2+k_c^2\right) }{k_b^2 k_c^2} P_L(k_b) P_L(k_c) \\ 
&\qquad + \text{2 permutations}  \,, \\
S_3(k_1,k_2) &= - \frac{1}{132} \int_{-1}^1 \text{d} \mu \, \frac{ \left(-k_a^2+k_b^2-k_c^2\right) \left(k_a^2+k_b^2-k_c^2\right) \left(-k_a^2+k_b^2+k_c^2\right)}{k_b^2 k_c^2}P_L(k_b) P_L(k_c)\\ 
&\qquad  + \text{2 permutations} \,, \numberthis 
\end{align*}
where $k_a=k_1$, $k_b=k_2$ and $k_c=|\vec{k}_1 - \vec{k}_2| = \sqrt{k_1^2 +k_2^2 - 2 \mu k_1 k_2}$, and the permutations considered are the three cyclic permutations of $\{ k_a, k_b, k_c \}$. For the covariance, we have
\begin{align*}
S_4(k_1,k_2) &=\frac{1}{5460 k_2^3 k_1^3} \bigg[ 30 k_2^7 k_1+274 k_2^5 k_1^3-110 k_2^3 k_1^5+30 k_2 k_1^7 \\
&\qquad +15 \left(k_2^2-k_1^2\right)^4 \log {|k_2 - k_1| \over k_2 + k_1 } \, \bigg] P_L(k_2) P_L(k_1)^2 + \left( k_1 \leftrightarrow k_2 \right)
\,, \\
S_5(k_1,k_2) &= \frac{1}{5460 k_2^3 k_1^5} \bigg[2  \left(15 k_2^9 k_1-40 k_2^7 k_1^3+82 k_2^5 k_1^5-40 k_2^3 k_1^7+15 k_2 k_1^9\right)  \\ 
&\qquad +  15 \left(k_2^2+k_1^2\right) \left(k_2^2-k_1^2\right)^4 \log { |k_2-k_1| \over k_2+k_1 }   \, \bigg] P_L(k_2) P_L(k_1)^2 + \left( k_1 \leftrightarrow k_2 \right)  \,, \\
S_6(k_1,k_2) &= -  \frac{6}{13}  k_2^2 P_L(k_2) P_L(k_1)^2 + \left( k_1 \leftrightarrow k_2 \right) \,. \numberthis
\end{align*}

\section{One-Loop Squeezed Bispectrum}
\label{app:OneLoopBs}
We collect diagrams and amplitudes in SPT and EFT for the one-loop contribution to the response function $G_1(k)$, defined in \Eq{eq:G1}.

\subsection{SPT}
\label{app:BsSPT}
The one-loop SPT diagrams 
and their corresponding expressions in the squeezed expansion, normalized by $P_L(k)P_L(q)$, are:
\begin{align*}
\begin{tikzpicture}[baseline={([yshift=-.5ex]current bounding box.center)}]
\draw[thick,dashed] (-0.5,0) -- (0,1);
\draw[ thick,dashed] (0.5,0) -- (0,1);
\draw [fill=black] (0,1) circle (2pt);
\draw [fill=black] (-0.5,0) circle (2pt);
\draw [fill=black] (0.5,0) circle (2pt);
\draw[thick,dashed] (0,1.39) ellipse (6pt and 12pt);
 \end{tikzpicture}  \mapsto
\frac{\langle B_{411}(\vec{k},\vec{k}',\vec{q})\rangle}{P_L(k)P_L(q)}&=\int
\frac{\df l\,P_L(l)}{42336\pi k^3 l^3}\Bigg\{8 k l(192 k^6-3419k^4l^2+2521k^2l^4-1230l^6)\\
&\quad+6(k^2-l^2)(64k^4+157k^2l^2-410l^4)\log\left(\frac{k-l}{k+l}\right)^2\\
&\quad-\frac{\df\log P_L(k)}{\df\log k}\Bigg[28(6k^7l-79k^5l^3+50k^3l^5-21kl^7)\\
&\quad+21(k^2-l^2)^3(2k^2+7l^2)\log\left(\frac{k-l}{k+l}\right)^2\Bigg]\Bigg\}+\mathcal{O}(q/k) \,, \numberthis\label{eq:B411}
\end{align*}
\begin{align*}
 \begin{tikzpicture}[baseline={([yshift=-.5ex]current bounding box.center)}]
\draw[ thick,dashed] (0.5,0) -- (0,1);
\draw [fill=black] (0,1) circle (2pt);
\draw [fill=black] (-0.5,0) circle (2pt);
\draw [fill=black] (0.5,0) circle (2pt);
\draw[rotate=-25,thick,dashed] (-0.45,0.35) ellipse (5pt and 15pt);
 \end{tikzpicture}  \mapsto 
\frac{\langle B_{321}^{a}(\vec{k},\vec{k}',\vec{q})\rangle}{P_L(k)P_L(q)}&=\int
\frac{\df l\df\nu\,P_L(l)k^4}{7056\pi^3P_L(k)}\Bigg\{\frac{28\pi k\nu+4\pi l(3-10\nu^2)}{(k^2+l^2-2kl\nu)^3}\Big[18k^3\nu-l^3(2-5\nu^2) \\
&\quad +k^2l(16-49\nu^2) -2kl^2\nu(19-25\nu^2)\Big]P_L\left(|\vec{k} - \vec{l} |\right)\\
&\quad+\frac{2\pi \left[7 k\nu- l(3-10\nu^2)\right]\left[141k\nu-l(59-200\nu^2)\right]}{(k^2+l^2+2kl\nu)^2} P_L\left(|\vec{k} + \vec{l} | \right) \\
&\quad-\frac{6\pi k(k-l\nu)\left[7k\nu+l(3-10\nu^2)\right]^2}{(k^2+l^2-2kl\nu)^{5/2}}P_L' \left(|\vec{k} - \vec{l} |\right) \Bigg\}+\mathcal{O}(q/k) \,, \numberthis\label{eq:B321a}
\end{align*}
\begin{align}\label{eq:B321b}
\begin{tikzpicture}[baseline={([yshift=-.5ex]current bounding box.center)}]
\draw[thick,dashed] (-0.5,0) -- (0,1);
\draw[ thick,dashed] (0.5,0) -- (0,1);
\draw [fill=black] (0,1) circle (2pt);
\draw [fill=black] (-0.5,0) circle (2pt);
\draw [fill=black] (0.5,0) circle (2pt);
\draw[rotate=45,thick,dashed] (-0.4,0.8) ellipse (6pt and 12pt);
 \end{tikzpicture}  \mapsto
\frac{\langle B_{321}^{b}(\vec{k},\vec{k}',\vec{q})\rangle}{P_L(k)P_L(q)}&=\frac{P_{13}(k)}{P_L(k)}\left[\frac{47}{42}-\frac{1}{6}\frac{\df \log P_{13}(k)}{\df\log k}\right]+\mathcal{O}(q/k) \, .
\end{align}
In Eq.~(\ref{eq:B321a}), $|\vec{k} \pm \vec{l} | = \sqrt{k^2 +l^2 \pm 2kl\nu }$ and $P_L'(k) = \df P_L(k)/\df k$, while in Eq.~(\ref{eq:B321b})
\begin{align}
P_{13}(k)=6P_L(k)\int \frac{\df^3 l\,P_L(l)}{(2\pi)^3}F_3^s(\vec{l},-\vec{l},\vec{k})
\end{align}
is the one-loop contribution to the power spectrum, and $F_3^s(\vec{l},-\vec{l},\vec{k})$ is the symmetrized SPT kernel. In the squeezed limit, $\langle B_\text{222}(\vec{k},\vec{k}',\vec{q})\rangle/\left[P_L(k)P_L(q)\right] = \mathcal{O}(q/k)$
and can thus be neglected at leading order.

In the expressions above, $\nu\in[-1,1]$ and $l\in[0,l_\text{max}]$. The UV cutoff $l_\text{max}$ is in principle arbitrary and
different choices would be reabsorbed into different values of the EFT counterterms. At a given order in the EFT expansion, the sum of the SPT loops and the counterterms is physical and thus independent of the cutoff up to that order. In our analysis we use $l_\text{max}=20~\text{h/Mpc}$, which is much larger than any scale of interest. 

Separate diagrams can be IR divergent, but Galilean invariance ensures that the sum is IR finite.  In particular, following Refs.~\cite{2015JCAP...05..007B,2014JCAP...07..056C}, one can define an IR-safe integrand by remapping the IR poles of $B_{321}^a$ at $\vec{l}\to\vec{k},\vec{k}'$ to poles at $\vec{l}\to 0$.  Once $B_{411}$ and $B_{321}^b$ are added, all the divergences are at $\vec{l}= 0$, and cancel by construction. The remapping involves phase space restrictions of the form $\Theta(|\vec{k}\pm\vec{l}|-l)$ or $\Theta(|\vec{k'}\pm\vec{l}|-l)$, and one should make sure to implement the remapping of the $B_{321}^a$ integrand before the squeezed expansion, as first-order terms from the Heaviside theta expansion also contribute at leading order. For our analysis, we compare to N-body simulation data from \cite{Wagner:2015gva}. In this case, we cut off the loop integrals at $l_\text{min}\equiv2\pi/L$, where $L$ is the size of the box in the simulation. Consistently, the linear power spectrum $P_L(k)$ is also cut off for $k<2\pi/L$.

\subsection{EFT}
\label{app:BsEFT}
The EFT diagrams and the corresponding amplitudes in the squeezed limit, normalized by $P_L(k)P_L(q)$, are:
\begin{align}
\begin{tikzpicture}[baseline={([yshift=-.5ex]current bounding box.center)}]
\draw[thick,dashed] (-0.5,0) -- (0,1);
\draw[ thick,dashed] (0.5,0) -- (0,1);
\draw [fill=lightgray] (-0.09,1.09) rectangle (0.09,0.91);
\draw [fill=black] (-0.5,0) circle (2pt);
\draw [fill=black] (0.5,0) circle (2pt);
 \end{tikzpicture} \mapsto
\frac{\langle \Delta B_{411}(\vec{k},\vec{k}',\vec{q})\rangle}{P_L(k)P_L(q)}=-\frac{8}{99}k^2(3\bar{c}_1+2\bar{c}_2+\bar{c}_3)-\bar{c}_sk^2\left(\frac{499}{2079}-\frac{k}{27}\frac{P_L'(k)}{P_L(k)}\right) + \order(q/k) ,
\end{align}
\begin{align}
\begin{tikzpicture}[baseline={([yshift=-.5ex]current bounding box.center)}]
\draw[thick,dashed] (-0.5,0) -- (0,1);
\draw[ thick,dashed] (0.5,0) -- (0,1);
\draw [fill=lightgray] (-0.41,0.09) rectangle (-0.59,-0.09);
\draw [fill=black] (0,1) circle (2pt);
\draw [fill=black] (0.5,0) circle (2pt);
 \end{tikzpicture} \mapsto
\frac{\langle \Delta B_{321}^{b}(\vec{k},\vec{k}',\vec{q})\rangle}{P_L(k)P_L(q)}=-\bar{c}_sk^2\left(\frac{11}{63}-\frac{k}{27}\frac{P_L'(k)}{P_L(k)}\right)+ \order(q/k) .
\end{align}
We denote these counterterm diagrams by $\Delta X$ where $X$ is the corresponding SPT diagram they renormalize. By summing these two contributions and using \Eq{eq:G1loop} with $P_\text{1-loop}(k)=-2/9\bar{c}_sk^2P_L(k)$, one gets $G_1^\text{EFT}(k)$ in \Eq{eq:G1eft}.

\subsection{VKPR Ansatz}
\label{app:VKPR}

The response $G_1(k)$ defined in \Eq{eq:G1} measures the change in the growth of the short modes, due to the long-wavelength isotropic background. An isotropic background $\delta_L$ can be reabsorbed into a modified cosmology, with a non-zero curvature (i.e., the separate-universe approach). The linear growth function can then be written as 
\begin{align}\label{eq:growth}
\bar{D}(\bar{a})=\left(1+\frac{13}{21}\delta_L\right)D(a) +\mathcal{O}(\delta_L^2),
\end{align} 
where the bar indicates quantities in the modified cosmology. Thus, at leading order
\begin{align}\label{eq:VKPRlin}
G_1^\text{tree}(k)=\frac{13}{21}\frac{\df\log P_L(k)}{\df\log D(a)}=\frac{26}{21}.
\end{align}
References~\cite{Valageas:2013zda,Kehagias:2013paa} extend this ansatz beyond tree-level, replacing $P_L(k)$ with $P(k)$ in \Eq{eq:VKPRlin}. This is equivalent to assuming that, at higher perturbative order, the growth of the short modes is just modified by higher powers of \Eq{eq:growth}.
The one-loop VKPR correction to the response is 
\begin{align}
G_\text{1,VKPR}^\text{1-loop}(k)=\frac{26}{21}\frac{P_\text{1-loop}(k)}{P_L(k)}.
\end{align}
It has been pointed out that, even though this ansatz is not exact beyond tree level, it gives a reasonable approximation for density correlators, with differences estimated to be a few percent~\cite{Ben-Dayan:2014hsa}. This agrees with our result (see \Fig{fig:G1}).

If we assume the VKPR ansatz, the counterterm depends only on the speed of sound $\bar{c}_s$,
\begin{align}
G_\text{1,VKPR}^\text{EFT}(k)=-\frac{52}{189}\bar{c}_sk^2 \,,
\end{align}
and thus we can use N-body data for the response to measure $\bar{c}_s$ (see Sec.~\ref{subsec:measure}).

\bibliographystyle{JHEP}
\bibliography{SqueezedLimits}

\begin{thebibliography}{10}
\providecommand*{\bibinfo}[2]{#2}
\providecommand*{\eprint}[1]{#1}
\providecommand*{\url}[1]{#1}
\bibitem{2005astro.ph.10346T}
\bibinfo{author}{{The Dark Energy Survey Collaboration}},
  \bibinfo{journal}{ArXiv Astrophysics e-prints}  (\bibinfo{date}{Oct. 2005}),
  \eprint{astro-ph/0510346}.
\bibitem{2013Msngr.154...44J}
\bibinfo{author}{J.~T.~A. {de Jong}}, \bibinfo{author}{K.~{Kuijken}},
  \bibinfo{author}{D.~{Applegate}}, \bibinfo{author}{K.~{Begeman}},
  \bibinfo{author}{A.~{Belikov}}, \bibinfo{author}{C.~{Blake}},
  \bibinfo{author}{J.~{Bout}}, \bibinfo{author}{D.~{Boxhoorn}},
  \bibinfo{author}{H.~{Buddelmeijer}}, \bibinfo{author}{A.~{Buddendiek}},
  \emph{et~al.}, \bibinfo{journal}{The Messenger}
  \bibinfo{volume}{\textbf{154}}, \bibinfo{pages}{44} (\bibinfo{date}{Dec.
  2013}).
\bibitem{2012SPIE.8446E..0ZM}
\bibinfo{author}{S.~{Miyazaki}}, \bibinfo{author}{Y.~{Komiyama}},
  \bibinfo{author}{H.~{Nakaya}}, \bibinfo{author}{Y.~{Kamata}},
  \bibinfo{author}{Y.~{Doi}}, \bibinfo{author}{T.~{Hamana}},
  \bibinfo{author}{H.~{Karoji}}, \bibinfo{author}{H.~{Furusawa}},
  \bibinfo{author}{S.~{Kawanomoto}}, \bibinfo{author}{T.~{Morokuma}},
  \emph{et~al.}, in \emph{Society of Photo-Optical Instrumentation Engineers
  (SPIE) Conference Series} (\bibinfo{date}{Sep. 2012}), \bibinfo{volume}{vol.
  8446 of \emph{Society of Photo-Optical Instrumentation Engineers (SPIE)
  Conference Series}}, \bibinfo{pages}{p.~0}.
\bibitem{2012arXiv1211.0310L}
\bibinfo{author}{{LSST Dark Energy Science Collaboration}},
  \bibinfo{journal}{ArXiv e-prints}  (\bibinfo{date}{Nov. 2012}),
  \eprint{1211.0310}.
\bibitem{2013AJ....145...10D}
\bibinfo{author}{K.~S. {Dawson}}, \bibinfo{author}{D.~J. {Schlegel}},
  \bibinfo{author}{C.~P. {Ahn}}, \bibinfo{author}{S.~F. {Anderson}},
  \bibinfo{author}{{\'E}.~{Aubourg}}, \bibinfo{author}{S.~{Bailey}},
  \bibinfo{author}{R.~H. {Barkhouser}}, \bibinfo{author}{J.~E. {Bautista}},
  \bibinfo{author}{A.~{Beifiori}}, \bibinfo{author}{A.~A. {Berlind}},
  \emph{et~al.}, \bibinfo{journal}{\aj} \bibinfo{volume}{\textbf{145}},
  \bibinfo{pages}{10}, \bibinfo{eid}{10} (\bibinfo{date}{Jan. 2013}),
  \eprint{1208.0022}.
\bibitem{2013arXiv1308.0847L}
\bibinfo{author}{M.~{Levi}}, \bibinfo{author}{C.~{Bebek}},
  \bibinfo{author}{T.~{Beers}}, \bibinfo{author}{R.~{Blum}},
  \bibinfo{author}{R.~{Cahn}}, \bibinfo{author}{D.~{Eisenstein}},
  \bibinfo{author}{B.~{Flaugher}}, \bibinfo{author}{K.~{Honscheid}},
  \bibinfo{author}{R.~{Kron}}, \bibinfo{author}{O.~{Lahav}}, \emph{et~al.},
  \bibinfo{journal}{ArXiv e-prints}  (\bibinfo{date}{Aug. 2013}),
  \eprint{1308.0847}.
\bibitem{2015arXiv150804473D}
\bibinfo{author}{K.~S. {Dawson}}, \bibinfo{author}{J.-P. {Kneib}},
  \bibinfo{author}{W.~J. {Percival}}, \bibinfo{author}{S.~{Alam}},
  \bibinfo{author}{F.~D. {Albareti}}, \bibinfo{author}{S.~F. {Anderson}},
  \bibinfo{author}{E.~{Armengaud}}, \bibinfo{author}{E.~{Aubourg}},
  \bibinfo{author}{S.~{Bailey}}, \bibinfo{author}{J.~E. {Bautista}},
  \emph{et~al.}, \bibinfo{journal}{ArXiv e-prints}  (\bibinfo{date}{Aug.
  2015}), \eprint{1508.04473}.
\bibitem{2011arXiv1106.1706S}
\bibinfo{author}{D.~{Schlegel}}, \bibinfo{author}{F.~{Abdalla}},
  \bibinfo{author}{T.~{Abraham}}, \bibinfo{author}{C.~{Ahn}},
  \bibinfo{author}{C.~{Allende Prieto}}, \bibinfo{author}{J.~{Annis}},
  \bibinfo{author}{E.~{Aubourg}}, \bibinfo{author}{M.~{Azzaro}},
  \bibinfo{author}{S.~B.~C. {Baltay}}, \bibinfo{author}{C.~{Baugh}},
  \emph{et~al.}, \bibinfo{journal}{ArXiv e-prints}  (\bibinfo{date}{Jun.
  2011}), \eprint{1106.1706}.
\bibitem{Ellis:2012rn}
\bibinfo{author}{R.~Ellis} \emph{et~al.} (\bibinfo{collaboration}{PFS Team}),
  \bibinfo{journal}{Publ. Astron. Soc. Jap.} \bibinfo{volume}{\textbf{66}}(1),
  \bibinfo{pages}{R1} (\bibinfo{date}{2014}), \eprint{1206.0737}.
\bibitem{2015arXiv150303757S}
\bibinfo{author}{D.~{Spergel}}, \bibinfo{author}{N.~{Gehrels}},
  \bibinfo{author}{C.~{Baltay}}, \bibinfo{author}{D.~{Bennett}},
  \bibinfo{author}{J.~{Breckinridge}}, \bibinfo{author}{M.~{Donahue}},
  \bibinfo{author}{A.~{Dressler}}, \bibinfo{author}{B.~S. {Gaudi}},
  \bibinfo{author}{T.~{Greene}}, \bibinfo{author}{O.~{Guyon}}, \emph{et~al.},
  \bibinfo{journal}{ArXiv e-prints}  (\bibinfo{date}{Mar. 2015}),
  \eprint{1503.03757}.
\bibitem{2013LRR....16....6A}
\bibinfo{author}{L.~{Amendola}}, \bibinfo{author}{S.~{Appleby}},
  \bibinfo{author}{D.~{Bacon}}, \bibinfo{author}{T.~{Baker}},
  \bibinfo{author}{M.~{Baldi}}, \bibinfo{author}{N.~{Bartolo}},
  \bibinfo{author}{A.~{Blanchard}}, \bibinfo{author}{C.~{Bonvin}},
  \bibinfo{author}{S.~{Borgani}}, \bibinfo{author}{E.~{Branchini}},
  \emph{et~al.}, \bibinfo{journal}{Living Reviews in Relativity}
  \bibinfo{volume}{\textbf{16}}, \bibinfo{pages}{6} (\bibinfo{date}{Sep.
  2013}), \eprint{1206.1225}.
\bibitem{2002PhR...367....1B}
\bibinfo{author}{F.~{Bernardeau}}, \bibinfo{author}{S.~{Colombi}},
  \bibinfo{author}{E.~{Gazta{\~n}aga}}, and \bibinfo{author}{R.~{Scoccimarro}},
  \bibinfo{journal}{\physrep} \bibinfo{volume}{\textbf{367}},
  \bibinfo{pages}{1} (\bibinfo{date}{Sep. 2002}), \eprint{astro-ph/0112551}.
\bibitem{2012JCAP...07..051B}
\bibinfo{author}{D.~{Baumann}}, \bibinfo{author}{A.~{Nicolis}},
  \bibinfo{author}{L.~{Senatore}}, and \bibinfo{author}{M.~{Zaldarriaga}},
  \bibinfo{journal}{\jcap} \bibinfo{volume}{\textbf{7}}, \bibinfo{pages}{51},
  \bibinfo{eid}{051} (\bibinfo{date}{Jul. 2012}), \eprint{1004.2488}.
\bibitem{2014PhRvD..89d3521H}
\bibinfo{author}{M.~P. {Hertzberg}}, \bibinfo{journal}{\prd}
  \bibinfo{volume}{\textbf{89}}(4), \bibinfo{pages}{043521},
  \bibinfo{eid}{043521} (\bibinfo{date}{Feb. 2014}), \eprint{1208.0839}.
\bibitem{2012JHEP...09..082C}
\bibinfo{author}{J.~J.~M. {Carrasco}}, \bibinfo{author}{M.~P. {Hertzberg}}, and
  \bibinfo{author}{L.~{Senatore}}, \bibinfo{journal}{Journal of High Energy
  Physics} \bibinfo{volume}{\textbf{9}}, \bibinfo{pages}{82}, \bibinfo{eid}{82}
  (\bibinfo{date}{Sep. 2012}), \eprint{1206.2926}.
\bibitem{Foreman:2015lca}
\bibinfo{author}{S.~Foreman}, \bibinfo{author}{H.~Perrier}, and
  \bibinfo{author}{L.~Senatore}, \bibinfo{journal}{JCAP}
  \bibinfo{volume}{\textbf{1605}}(05), \bibinfo{pages}{027}
  (\bibinfo{date}{2016}), \eprint{1507.05326}.
\bibitem{2016arXiv160200674B}
\bibinfo{author}{T.~{Baldauf}}, \bibinfo{author}{M.~{Mirbabayi}},
  \bibinfo{author}{M.~{Simonović}}, and \bibinfo{author}{M.~{Zaldarriaga}},
  \bibinfo{journal}{ArXiv e-prints}  (\bibinfo{date}{Feb. 2016}),
  \eprint{1602.00674}.
\bibitem{Bertolini:2016bmt}
\bibinfo{author}{D.~Bertolini}, \bibinfo{author}{K.~Schutz},
  \bibinfo{author}{M.~P. Solon}, and \bibinfo{author}{K.~M. Zurek},
  \bibinfo{journal}{JCAP} \bibinfo{volume}{\textbf{1606}}(06),
  \bibinfo{pages}{052} (\bibinfo{date}{2016}), \eprint{1604.01770}.
\bibitem{2015JCAP...05..007B}
\bibinfo{author}{T.~{Baldauf}}, \bibinfo{author}{L.~{Mercolli}},
  \bibinfo{author}{M.~{Mirbabayi}}, and \bibinfo{author}{E.~{Pajer}},
  \bibinfo{journal}{\jcap} \bibinfo{volume}{\textbf{5}}, \bibinfo{pages}{7},
  \bibinfo{eid}{007} (\bibinfo{date}{May 2015}), \eprint{1406.4135}.
\bibitem{2015arXiv151207630B}
\bibinfo{author}{D.~Bertolini}, \bibinfo{author}{K.~Schutz},
  \bibinfo{author}{M.~P. Solon}, \bibinfo{author}{J.~R. Walsh}, and
  \bibinfo{author}{K.~M. Zurek}, \bibinfo{journal}{Phys. Rev.}
  \bibinfo{volume}{\textbf{D93}}(12), \bibinfo{pages}{123505}
  (\bibinfo{date}{2016}), \eprint{1512.07630}.
\bibitem{Welling:2016dng}
\bibinfo{author}{Y.~{Welling}}, \bibinfo{author}{D.~{van der Woude}}, and
  \bibinfo{author}{E.~{Pajer}}, \bibinfo{journal}{ArXiv e-prints}
  (\bibinfo{date}{May 2016}), \eprint{1605.06426}.
\bibitem{Kehagias:2013yd}
\bibinfo{author}{A.~Kehagias} and \bibinfo{author}{A.~Riotto},
  \bibinfo{journal}{Nucl. Phys.} \bibinfo{volume}{\textbf{B873}},
  \bibinfo{pages}{514} (\bibinfo{date}{2013}), \eprint{1302.0130}.
\bibitem{Peloso:2013zw}
\bibinfo{author}{M.~Peloso} and \bibinfo{author}{M.~Pietroni},
  \bibinfo{journal}{JCAP} \bibinfo{volume}{\textbf{1305}}, \bibinfo{pages}{031}
  (\bibinfo{date}{2013}), \eprint{1302.0223}.
\bibitem{Creminelli:2013mca}
\bibinfo{author}{P.~Creminelli}, \bibinfo{author}{J.~Noreña},
  \bibinfo{author}{M.~Simonović}, and \bibinfo{author}{F.~Vernizzi},
  \bibinfo{journal}{JCAP} \bibinfo{volume}{\textbf{1312}}, \bibinfo{pages}{025}
  (\bibinfo{date}{2013}), \eprint{1309.3557}.
\bibitem{Peloso:2013spa}
\bibinfo{author}{M.~Peloso} and \bibinfo{author}{M.~Pietroni},
  \bibinfo{journal}{JCAP} \bibinfo{volume}{\textbf{1404}}, \bibinfo{pages}{011}
  (\bibinfo{date}{2014}), \eprint{1310.7915}.
\bibitem{Creminelli:2013poa}
\bibinfo{author}{P.~Creminelli}, \bibinfo{author}{J.~Gleyzes},
  \bibinfo{author}{M.~Simonović}, and \bibinfo{author}{F.~Vernizzi},
  \bibinfo{journal}{JCAP} \bibinfo{volume}{\textbf{1402}}, \bibinfo{pages}{051}
  (\bibinfo{date}{2014}), \eprint{1311.0290}.
\bibitem{Kehagias:2013rpa}
\bibinfo{author}{A.~Kehagias}, \bibinfo{author}{J.~Noreña},
  \bibinfo{author}{H.~Perrier}, and \bibinfo{author}{A.~Riotto},
  \bibinfo{journal}{Nucl. Phys.} \bibinfo{volume}{\textbf{B883}},
  \bibinfo{pages}{83} (\bibinfo{date}{2014}), \eprint{1311.0786}.
\bibitem{Valageas:2013cma}
\bibinfo{author}{P.~Valageas}, \bibinfo{journal}{Phys. Rev.}
  \bibinfo{volume}{\textbf{D89}}(8), \bibinfo{pages}{083534}
  (\bibinfo{date}{2014}), \eprint{1311.1236}.
\bibitem{Valageas:2013zda}
\bibinfo{author}{P.~Valageas}, \bibinfo{journal}{Phys. Rev.}
  \bibinfo{volume}{\textbf{D89}}(12), \bibinfo{pages}{123522}
  (\bibinfo{date}{2014}), \eprint{1311.4286}.
\bibitem{Kehagias:2013paa}
\bibinfo{author}{A.~Kehagias}, \bibinfo{author}{H.~Perrier}, and
  \bibinfo{author}{A.~Riotto}, \bibinfo{journal}{Mod. Phys. Lett.}
  \bibinfo{volume}{\textbf{A29}}, \bibinfo{pages}{1450152}
  (\bibinfo{date}{2014}), \eprint{1311.5524}.
\bibitem{Creminelli:2013nua}
\bibinfo{author}{P.~Creminelli}, \bibinfo{author}{J.~Gleyzes},
  \bibinfo{author}{L.~Hui}, \bibinfo{author}{M.~Simonović}, and
  \bibinfo{author}{F.~Vernizzi}, \bibinfo{journal}{JCAP}
  \bibinfo{volume}{\textbf{1406}}, \bibinfo{pages}{009} (\bibinfo{date}{2014}),
  \eprint{1312.6074}.
\bibitem{Nishimichi:2014jna}
\bibinfo{author}{T.~Nishimichi} and \bibinfo{author}{P.~Valageas},
  \bibinfo{journal}{Phys. Rev.} \bibinfo{volume}{\textbf{D90}}(2),
  \bibinfo{pages}{023546} (\bibinfo{date}{2014}), \eprint{1402.3293}.
\bibitem{Ben-Dayan:2014hsa}
\bibinfo{author}{I.~Ben-Dayan}, \bibinfo{author}{T.~Konstandin},
  \bibinfo{author}{R.~A. Porto}, and \bibinfo{author}{L.~Sagunski},
  \bibinfo{journal}{JCAP} \bibinfo{volume}{\textbf{1502}}(02),
  \bibinfo{pages}{026} (\bibinfo{date}{2015}), \eprint{1411.3225}.
\bibitem{Horn:2015dra}
\bibinfo{author}{B.~Horn}, \bibinfo{author}{L.~Hui}, and
  \bibinfo{author}{X.~Xiao}, \bibinfo{journal}{JCAP}
  \bibinfo{volume}{\textbf{1509}}(09), \bibinfo{pages}{068}
  (\bibinfo{date}{2015}), \eprint{1502.06980}.
\bibitem{Dai:2015jaa}
\bibinfo{author}{L.~Dai}, \bibinfo{author}{E.~Pajer}, and
  \bibinfo{author}{F.~Schmidt}, \bibinfo{journal}{JCAP}
  \bibinfo{volume}{\textbf{1510}}(10), \bibinfo{pages}{059}
  (\bibinfo{date}{2015}), \eprint{1504.00351}.
\bibitem{Sherwin:2012nh}
\bibinfo{author}{B.~D. Sherwin} and \bibinfo{author}{M.~Zaldarriaga},
  \bibinfo{journal}{Phys. Rev.} \bibinfo{volume}{\textbf{D85}},
  \bibinfo{pages}{103523} (\bibinfo{date}{2012}), \eprint{1202.3998}.
\bibitem{Takada:2013bfn}
\bibinfo{author}{M.~Takada} and \bibinfo{author}{W.~Hu},
  \bibinfo{journal}{Phys. Rev.} \bibinfo{volume}{\textbf{D87}}(12),
  \bibinfo{pages}{123504} (\bibinfo{date}{2013}), \eprint{1302.6994}.
\bibitem{Li:2014sga}
\bibinfo{author}{Y.~Li}, \bibinfo{author}{W.~Hu}, and
  \bibinfo{author}{M.~Takada}, \bibinfo{journal}{Phys. Rev.}
  \bibinfo{volume}{\textbf{D89}}(8), \bibinfo{pages}{083519}
  (\bibinfo{date}{2014}), \eprint{1401.0385}.
\bibitem{Wagner:2015gva}
\bibinfo{author}{C.~Wagner}, \bibinfo{author}{F.~Schmidt},
  \bibinfo{author}{C.-T. Chiang}, and \bibinfo{author}{E.~Komatsu},
  \bibinfo{journal}{JCAP} \bibinfo{volume}{\textbf{1508}}(08),
  \bibinfo{pages}{042} (\bibinfo{date}{2015}), \eprint{1503.03487}.
\bibitem{Wagner:2014aka}
\bibinfo{author}{C.~Wagner}, \bibinfo{author}{F.~Schmidt},
  \bibinfo{author}{C.-T. Chiang}, and \bibinfo{author}{E.~Komatsu},
  \bibinfo{journal}{Mon. Not. Roy. Astron. Soc.}
  \bibinfo{volume}{\textbf{448}}(1), \bibinfo{pages}{L11}
  (\bibinfo{date}{2015}), \eprint{1409.6294}.
\bibitem{Pajer:2013jj}
\bibinfo{author}{E.~Pajer} and \bibinfo{author}{M.~Zaldarriaga},
  \bibinfo{journal}{JCAP} \bibinfo{volume}{\textbf{1308}}, \bibinfo{pages}{037}
  (\bibinfo{date}{2013}), \eprint{1301.7182}.
\bibitem{2014arXiv1406.4143A}
\bibinfo{author}{R.~E. {Angulo}}, \bibinfo{author}{S.~{Foreman}},
  \bibinfo{author}{M.~{Schmittfull}}, and \bibinfo{author}{L.~{Senatore}},
  \bibinfo{journal}{ArXiv e-prints}  (\bibinfo{date}{Jun. 2014}),
  \eprint{1406.4143}.
\bibitem{2014JCAP...07..057C}
\bibinfo{author}{J.~J.~M. {Carrasco}}, \bibinfo{author}{S.~{Foreman}},
  \bibinfo{author}{D.~{Green}}, and \bibinfo{author}{L.~{Senatore}},
  \bibinfo{journal}{\jcap} \bibinfo{volume}{\textbf{7}}, \bibinfo{pages}{57},
  \bibinfo{eid}{057} (\bibinfo{date}{Jul. 2014}), \eprint{1310.0464}.
\bibitem{Baldauf:2015aha}
\bibinfo{author}{T.~Baldauf}, \bibinfo{author}{L.~Mercolli}, and
  \bibinfo{author}{M.~Zaldarriaga}, \bibinfo{journal}{Phys. Rev.}
  \bibinfo{volume}{\textbf{D92}}(12), \bibinfo{pages}{123007}
  (\bibinfo{date}{2015}), \eprint{1507.02256}.
\bibitem{Baldauf:2015zga}
\bibinfo{author}{T.~Baldauf}, \bibinfo{author}{E.~Schaan}, and
  \bibinfo{author}{M.~Zaldarriaga}, \bibinfo{journal}{JCAP}
  \bibinfo{volume}{\textbf{1603}}(03), \bibinfo{pages}{007}
  (\bibinfo{date}{2016}), \eprint{1507.02255}.
\bibitem{Dai:2015rda}
\bibinfo{author}{L.~Dai}, \bibinfo{author}{E.~Pajer}, and
  \bibinfo{author}{F.~Schmidt}, \bibinfo{journal}{JCAP}
  \bibinfo{volume}{\textbf{1511}}(11), \bibinfo{pages}{043}
  (\bibinfo{date}{2015}), \eprint{1502.02011}.
\bibitem{Garny:2015oya}
\bibinfo{author}{M.~Garny}, \bibinfo{author}{T.~Konstandin},
  \bibinfo{author}{R.~A. Porto}, and \bibinfo{author}{L.~Sagunski},
  \bibinfo{journal}{JCAP} \bibinfo{volume}{\textbf{1511}}(11),
  \bibinfo{pages}{032} (\bibinfo{date}{2015}), \eprint{1508.06306}.
\bibitem{1986ApJ...311....6G}
\bibinfo{author}{M.~H. {Goroff}}, \bibinfo{author}{B.~{Grinstein}},
  \bibinfo{author}{S.-J. {Rey}}, and \bibinfo{author}{M.~B. {Wise}},
  \bibinfo{journal}{\apj} \bibinfo{volume}{\textbf{311}}, \bibinfo{pages}{6}
  (\bibinfo{date}{Dec. 1986}).
\bibitem{1996ApJS..105...37S}
\bibinfo{author}{R.~{Scoccimarro}} and \bibinfo{author}{J.~{Frieman}},
  \bibinfo{journal}{\apjs} \bibinfo{volume}{\textbf{105}}, \bibinfo{pages}{37}
  (\bibinfo{date}{Jul. 1996}), \eprint{astro-ph/9509047}.
\bibitem{2014JCAP...07..056C}
\bibinfo{author}{J.~J.~M. {Carrasco}}, \bibinfo{author}{S.~{Foreman}},
  \bibinfo{author}{D.~{Green}}, and \bibinfo{author}{L.~{Senatore}},
  \bibinfo{journal}{\jcap} \bibinfo{volume}{\textbf{7}}, \bibinfo{pages}{56},
  \bibinfo{eid}{056} (\bibinfo{date}{Jul. 2014}), \eprint{1304.4946}.

\end{thebibliography}
\end{document}